\documentclass{ieeeaccess}
\usepackage{cite}
\usepackage{amsmath,amssymb,amsfonts}
\usepackage{algorithmic}
\usepackage{graphicx}
\usepackage{textcomp}
\usepackage{subcaption}

\usepackage{booktabs}
\usepackage{multirow}
\usepackage{color}

\newcommand{\civ}{{\sc civ}}

\newcommand{\access}[1]{\mathit{access(#1)}}
\newcommand{\next}[1]{\mathit{next(#1)}}

\def\BibTeX{{\rm B\kern-.05em{\sc i\kern-.025em b}\kern-.08em
    T\kern-.1667em\lower.7ex\hbox{E}\kern-.125emX}}
\begin{document}
\history{Received September 27, 2019, accepted October 20, 2019, date of publication October 25, 2019, date of current version November 6, 2019.}
\doi{10.1109/ACCESS.2019.2949655}

\title{Energy consumption in compact integer vectors: A study case}
\author{\uppercase{Jos{\'e} Fuentes-Sep{\'u}lveda}\authorrefmark{1},
\uppercase{Susana Ladra}\authorrefmark{2}}
\address[1]{Universidad de Chile, Department of Computer Science, Santiago, Chile (e-mail: jfuentess@dcc.uchile.cl)}
\address[2]{Universidade da Coru{\~n}a, CITIC, Facultad de Inform{\'a}tica, A Coru{\~n}a, Spain (e-mail: susana.ladra@udc.es)}
\tfootnote{This research has received funding from the European Union's Horizon 2020 research and innovation programme under the Marie Sklodowska-Curie [grant agreement No 690941]; from the Ministerio de Ciencia, Innovaci{\'o}n y Universidades (PGE and ERDF) [grant numbers TIN2016-77158-C4-3-R; RTC-2017-5908-7]; from
Xunta de Galicia (co-founded with ERDF) [grant numbers ED431C 2017/58; ED431G/01]; from postdoctoral grant of Comisi\'on Nacional de Investigaci\'on Cient\'ifica y Tecnol\'ogica, CONICYT [grant number 3170534].}

\markboth
{Fuentes-Sep{\'u}lveda \headeretal: Energy consumption in compact integervectors: A study case}
{Fuentes-Sep{\'u}lveda \headeretal: Energy consumption in compact integervectors: A study case}

\corresp{Corresponding author: Susana Ladra (e-mail: susana.ladra@udc.es).}

\begin{abstract}
In the field of algorithms and data structures analysis and design, most of the researchers focus only on the space/time trade-off, and little attention has been paid to energy consumption. Moreover, most of the efforts in the field of Green Computing have been devoted to hardware-related issues, being green software in its infancy.
Optimizing the usage of computing resources, minimizing power consumption or increasing battery life are some of the goals of this field of research.

As an attempt to address the most recent sustainability challenges, we must incorporate the energy consumption as a first-class constraint when designing new compact data structures. Thus, as a preliminary work to reach that goal, we first need to understand the factors that impact on the energy consumption and their relation with compression. In this work, we study the energy consumption required by several integer vector representations. We execute typical operations over datasets of different nature. We can see that, as commonly believed, energy consumption is highly related to the time required by the process, but not always. We analyze other parameters, such as number of instructions, number of CPU cycles, memory loads, among others.
\end{abstract}


\titlepgskip=-15pt

\maketitle

\section{Introduction}

We are surrounded by digital information, such as the huge amount of data
generated on the Internet and also that we are collecting in our daily lives:
human generated data, both consciously (such as emails, tweets, pictures, voice)
and unconsciously (clicks, likes, follows, logs, ...), or observed data
(biological, astronomical, etc.). When managing large volumes of digital
information, data compression has always been considered to be vitally
important. Traditionally, data compression focused on obtaining the smallest
representation possible, in order to save space and transmission time, thus,
providing a good archival method. However, most of the compression techniques
require decompressing the data when they need to be accessed, especially when
these accesses are not sequential, and thus, limiting the applicability of data
compression. 

To overcome these issues, {\em compact data structures} appeared in the 1990's and
rapidly evolved during early years of the current century~\cite{Nav16}. They
use compression strategies to reduce the size of the stored data, taking
advantage of the patterns existing in the data, but with a key difference: data
can be directly managed and queried in compressed form, without requiring prior decompression. The main contribution is that they allow larger datasets
fit in faster levels of the memory hierarchy than classical representations,
thus, dramatically improving processing times. In addition, many compact data
structures are equipped with additional information that, within the same
compressed space, acts as index and speeds up queries.  

Nowadays, multiple compact data structures have been proposed for representing
data of different nature, from simple bitvectors or sequences, to other complex
types of data, such as permutations, trees, grids, binary relations, graphs,
tries, text collections, and others~\cite{Nav16}. Compact data structures
have now reached a maturity level, with some of them being used in real systems
from other communities, such as in Information Retrieval or Bioinformatics. For
instance, some of these compressed data structures are the core of the existing
genome assemblers or DNA aligners~\cite{Langmead2009,10.1093/bioinformatics/btp324}, or have been shown competitive compared with traditional solutions for text or document
retrieval~\cite{Culpepper:2010:TRD:1882123.1882145,Konow:2012}. 

Despite of the benefits that compact data structures can provide, they are not
present in most of the internet-connected devices that are part of our daily
lives. There are now more of these devices than there are people in the world,
and is expected that by 2020 they will outnumber humans by 4-to-1, based on the
last forecast by Gartner. Smartphones, smart TVs, wearables, healthcare sensors,
etc., are expected to proliferate in the near future, as the development of
smart cities, industry 4.0, autonomous cars or unmanned aerial vehicles is
becoming a reality. Most of these devices include decision-making algorithms
that require collecting, storing and processing large amounts of data using
little space, which would seem to be a perfect scenario for compact data structures. Compact data structures have received few attention in these application domains, and this can be related to disregarding energy at the design step of compact data structures. Addressing one of the main biggest limitation of these devices, their battery life, is vital, as they generally
operate far from any power supply. Up to today, few researchers in the field of
compact data structures expressed any concern about energy
consumption~\cite{10.1007/978-3-642-15781-3_1}, and there is no previous work on
designing energy-aware compact data structures.  

Power management and energy efficiency have been the main focus in Green
Computing. Most of the research efforts in this field were devoted to reduce
data centers' carbon footprint, or dealing with electrical and computer systems
engineering~\cite{VEIGA2018565}. However, we are still far from being able to
develop energy-efficient software due to two main problems: the lack of
knowledge and the lack of tools~\cite{Pinto:2017:EEN:3167461.3154384}. Despite this fact,
green software is gaining importance, especially in the areas of quality and
design/construction, followed by requirements~\cite{CALERO2017117}. There
are also some previous research on rethinking the way query processing was
performed to become energy-aware, such as in the case of database management
systems~\cite{Lang2011RethinkingQP,10.1007/978-3-642-55149-9_4}.  
Nevertheless, there is an absence of any science of power management, with a
relatively small amount of research in the core areas of computer science~\cite{5233517,Ranganathan:2010:REP:1721654.1721673}. 

To date, compressing data has been considered before as a technique to reduce
energy requirements~\cite{Sadler:2006:DCA:1182807.1182834}, as it can improve the way data
is stored in memory, moving them as close as possible to the processing
entity~\cite{Larsson2008}. However, the design of energy-efficient compact data
structures 
is challenging, as compact data structures usually rearrange data to enable fast
accesses while being compressed. These rearrangements would probably lead to
memory pattern accesses such those that negatively impact energy
consumption~\cite{Roy1997} and even increase temperature
~\cite{5599079}. Moreover, accessing compact data structures generally requires
more complex computations than their plain counterparts, thus, also increasing energy
requirements~\cite{Larsson2008}. In addition, not all compact data structures allow
increasing simultaneously time and space performance~\cite{Nav16}, thus,
energy comes into play as a new dimension to be taken into account. 

In this paper, we present a case of study in compact integer vectors, trying to get a deeper understanding of the different parameters and factors that impact on energy consumption. 

\section{Background}

\subsection{Background on Compressed Data Structures}
Data compression has always been a popular research area, as the amount of data we manage has been increasing continuously beyond the capabilities of the storage and processing systems \cite{Chen:2014}. Compression is not only a matter of reducing space disk, but most importantly, reducing transmission and processing times \cite{Salomon2008}. Thus, as we are now dealing with huge volumes of data of different nature, compression is definitively presented as a real solution to address, in parallel with other approaches, the phenomenon of what is commonly known as Big Data.  

Traditional compression techniques are usually designed to achieve the highest compression rate possible. However, there exist numerous scenarios where it is desirable to obtain better processing capabilities over the data than just obtain the most reduced space possible. For instance, there exist text compression techniques that allow for the retrieval of snippets located at random positions in the text without the need for decompressing the whole text \cite{Turpin:1997,Moura:2000}, which is a very desirable property in many applications, as compression and decompression phases are demanding processes both in terms of memory and time consumption. Moreover, it has been proven that searching string patterns directly over the compressed text is up to 8 times faster than searching over plain text \cite{Moura:2000,Brisaboa:2007}. Of course, these capabilities are usually provided at the expense of some compression ratio. 

In addition to the data itself, indexes or additional structures are usually required for supporting processing or querying over the data in efficient time. These data structures can be one or even two orders of magnitude larger than the data \cite{Nav16}. One of the classic examples is the suffix tree that enables efficient sequence analysis over one genome \cite{Weiner:1973}. In the case of human genomes, the data itself can occupy no more than 800 megabytes, whereas the suffix tree requires more than 30 gigabytes. Compact data structures are precisely aimed at addressing these issues. They maintain the data and their additional indexes using less space than the data itself and allowing efficient processing and querying of the data, without the need of decompression. 

The beginnings of compact data structures can be dated back to the year 1988, when Jacobson introduced, in his PhD thesis, a new set of data structures, named succinct data structures, which use $\log N + o(\log N)$ bits for representing $N$ different objects \cite{Jacobson1989}\footnote{We
use $\log x$ to mean the base 2 logarithm of $x$ unless specified otherwise.}. Some authors also used the term compressed data structure when proposing new data structures that use $\mathcal{H} + o(\log N)$, $\mathcal{H}$ being the entropy of the data under some compression model \cite{Ferragina:2005,Gupta:2007,Raman2015}. In the last few years, several other proposals have emerged, boosting the maturity of this field. The term {\it compact data structures} has recently been adopted by the community \cite{Nav16} to include all these types of data structures, generalizing the term to denote all data structures that use little space and query time, without specifying explicit complexity bounds. There exist multiple compact data structures for different problems: from the most basic needs such as representing arrays supporting reading and writing values at arbitrary positions \cite{Jacobson1989a,Raman:2007,FERRAGINA2007115}, bitvectors supporting bit-counting operations \cite{Okanohara2007,Claude2008,Gog:2014}, permutations \cite{Munro:2012,Barbay:2013}, sequences of symbols supporting counting and searching operations \cite{Grossi2003,Makinen:2005,Barbay:2014a,Belazzougui:2015}, to more complex data, such as trees \cite{Jacobson1989a,RamanRao:2013,NavarroSadakane:2014}, graphs \cite{Farzan:2013,Li2019}, grids \cite{Chan2011,Barbay:2013a,Brisaboa:2014}, texts \cite{Arroyuelo:2010,Culpepper:2010:TRD:1882123.1882145,Konow:2012}, geographical information \cite{deBernardo:2013,Ladra:2017}, etc.

\subsection{Background on Green Computing}

One of the main challenges that information technologies have to face is to
strike a balance between their enormous potential for growth and the
environmental impact that this is causing to our planet. There is an increasing
awareness by the various actors (researchers, governments, industry, etc.) of
the importance of developing technological solutions that make an efficient use
of energy and other resources, thereby promoting a long-term technological
sustainability. Energy savings in data centers, ecological manufacturing of
components, extending equipment longevity, etc. are some of the goals of what
has been called Green Computing. 

Much effort in Green Computing has been devoted to promoting a wise usage of the
resources and improving energy consumptions of hardware components or IT
infrastructures. However, the industry road map needs to move its efforts to
finding solutions in Green Software~\cite{DBLP:journals/nature/Waldrop16}. Instead of making better
chips, the focus will be centered in the different applications developed for
smartphones, supercomputers, or even data centers in the cloud. Then, after
understanding what these efficient high-level applications require, it will be
possible to go down to develop the chips and hardware to support their
needs. Research in Green Software is therefore gaining more attention,
especially in the areas of quality and design/construction, followed by
requirements~\cite{CALERO2017117}.  
Moreover, computing is not defined any more by the needs of traditional PCs or
data centers. Today, many people live surrounded by multiple mobile computing
devices, such as smartphones, tablets, wearables, smart TVs, smart home
products, and other kinds of sensors and IoT devices. In this scenario, energy
efficient software is of extreme importance, as it can save battery and reduce
the heat generated in the devices, which in turns increases the speed and
longevity of these mobile devices. Green IoT has emerged as a field of research
to tackle this problem, proposing generally hardware-based approaches, in
addition to other efforts that can be categorized as software-based,
habitual-based, awareness-based, policy-based, or
recycling-based~\cite{7997698}; however, approaches using compact data structure
have not been used in this context up to now. 

Among the previous work related to energy efficiency, there is little research
within the core areas of computer science~\cite{5233517}. Most of this research
focuses on online problems related to power management, including energy saving
mechanisms based on power-down mechanisms and speed
scaling~\cite{Albers:2010:EA:1735223.1735245}. Regarding energy-efficient
algorithm design, some authors have proposed 
new approaches, such as the reversible computation approach, where inputs can be
reconstructed from their outputs~\cite{Demaine:2016:EA:2840728.2840756}. Besides
this research, some works exist focusing on finding models or understanding the
factors affecting energy
consumption~\cite{Jain:2005:TUA:1080810.1080823,Roy:2013:ECM:2422436.2422470}. However, 
there is no research in the particular field of compact data structures; thus,
it becomes crucial to study the relationship between the spatial and temporal
complexities with the energy dimension including factors such as entropy, among
others. 

To date, there is a recognized lack of knowledge and tools for the development
of energy-efficient software~\cite{Pinto:2017:EEN:3167461.3154384}, which must firstly be
tackled to properly address the ecodesign of compact data structures. This
absence of any science of power management was identified almost 10 years ago~\cite{5233517,Ranganathan:2010:REP:1721654.1721673}, and there have been no significant advance
since then in the field of algorithms and data structures analysis and design.  
There exist some models proposed for database management
systems, software energy optimization for embedded systems~\cite{Lang2011RethinkingQP,10.1007/978-3-642-55149-9_4,Chatterjee2017}, or more general energy models
for algorithmic engineering~\cite{Jain:2005:TUA:1080810.1080823,Roy:2013:ECM:2422436.2422470}. 

\subsection{Greening Compact Data Structures: the Challenges}

The space reduction achieved by compact data structures usually comes at the expense of degrading the locality of reference, and this can significantly damage energy consumption \cite{Mussol:1997}. As an example, one can consider wavelet trees \cite{Grossi2003}, one of the most known and used compact data structures thanks to its multiple applications in the field of information retrieval \cite{Navarro:2014}. Wavelet trees are data structures that represent sequences of symbols in little space and can answer some queries over them efficiently. As most compact data structures, wavelet trees rearrange data in order to obtain good spatial and temporal properties, in addition to enabling fast counting, searching and direct accessing over the sequence. These rearrangements cause a lack of locality of reference, and consequently, memory access patterns that degrade energy consumption.

More concretely, given a sequence $S$ of length $n$, $S = S_1 S_2 \dots S_n$, composed of symbols from an alphabet $\Sigma = \{s_1, s_2, \dots, s_\sigma\}$, the wavelet tree is a balanced binary tree that divides the alphabet into two parts at each node. Each of these nodes of the tree contains a bitvector, which represents, for each position, if the corresponding symbol belongs to the lower (0) or to the upper (1) partition of the alphabet. More concretely, the recursive construction of the wavelet tree is as follows. The root node of the tree contains a bitvector ($B= b_1 b_2 … b_n$) of the same size as the sequence ($n$), where at position $i$ we have $b_i =0$ in case $S_i \in \{s_1, s_2, \dots, s_{\sigma/2}\}$, or $b_i =1$ in case $S_i \in \{ s_{\sigma/2+1},\dots, s_\sigma\}$. Those symbols marked with a 0 are processed in the left child of the node, and those marked 1 are processed in the right child of the node. Then, for each internal node the procedure is repeated recursively. As the alphabet indexed by a child node is only half of that of its parent, the construction stops when the alphabet is composed of just one symbol.

\begin{figure}[t]
  \centering
    \includegraphics[width=0.985\linewidth]{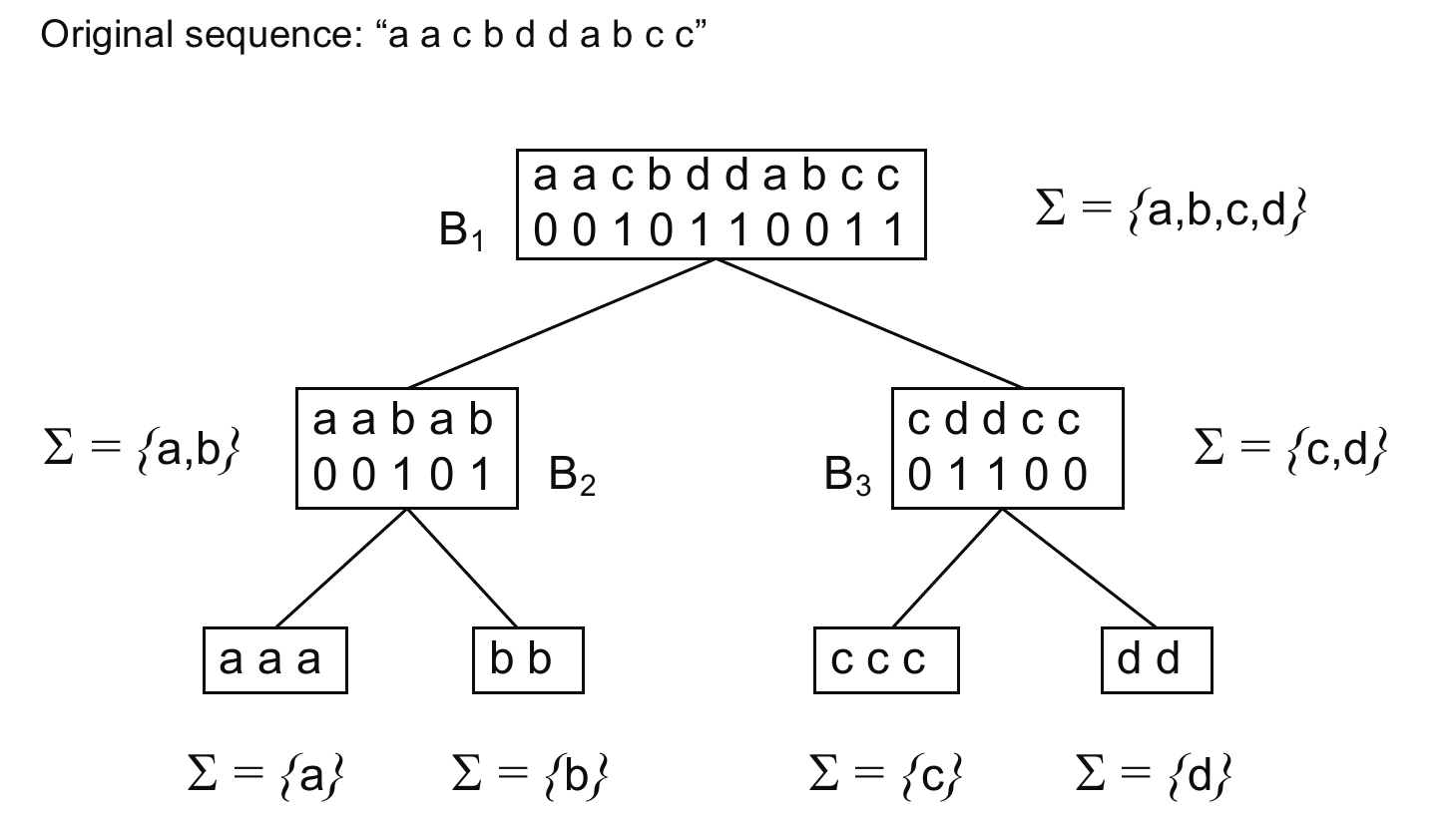}
  \caption{Example with a sequence of symbols from the alphabet $\Sigma = \{a, b, c, d\}$ and its corresponding wavelet tree. Text is shown only for clarity, but is not actually stored.}
  \label{fig:wt}
\end{figure}

Figure \ref{fig:wt} shows an example of a wavelet tree for the sequence ``$aacbddabcc$'' over the alphabet $\Sigma=\{a,b,c,d\}$. Only the bitmaps at each internal node are necessary to represent the sequence, as the alphabet corresponding to each node can be implicitly recovered following the path from the root tree to that node. Due to its construction procedure, the wavelet tree has a leaf node for each symbol of the alphabet. 

As can be seen in Figure \ref{fig:wt}, a sequence represented in plain form ($aacbddabcc$), is now rearranged in three different bitmaps following a two-level tree. Thanks to these rearrangements, this representation provides multiple benefits, as it allows counting and searching symbols in the sequence in efficient time and within less space. The plain form is only superior in terms of speed when recovering sequentially some substring of the sequence. As the size of the alphabet grows, the number of levels of the tree increases, and thus, extracting a portion of the original sequence requires multiple accesses to several nodes of the wavelet tree that are not contiguous in main memory, therefore, causing cache misses, movements of data within the memory hierarchy, generating a high energy consumption.

These rearrangements are also present in compact representation of trees, graphs, binary relations and texts. An example of this is the Burrows-Wheeler Transform (BWT) \cite{Burrows94}, a classical algorithm in the field of data compression, which is used, for instance, in the known compressor bzip2. Given a string of characters, BWT rearranges them with the goal of obtaining long runs of similar characters, improving its compressibility. The BWT is the basis of the FM-Index \cite{Ferragina:2000}, which is at the core of well-known bioinformatics software, such as the Burrows-Wheeler Aligner (BWA) \cite{10.1093/bioinformatics/btp324}, Bowtie aligner \cite{Langmead2009}, or the SGA Assembler \cite{Simpson:2010}. The FM-index uses a technique for locating patterns in the text called backwards search, which jumps to random positions of the rearranged text to find the occurrences of the searched pattern. Again, this may cause cache misses, movements of data within the memory hierarchy, which penalize energy consumption.

\section{Study Case: Compressed Integer Vectors}

\subsection{Compressed Integer Vectors}

\begin{table*}[t]
\begin{center}
  \begin{tabular}{l@{\hspace{1ex}}c@{\hspace{1ex}}c@{\hspace{1ex}}c@{\hspace{1ex}}l}
    \toprule
    & Space (bits) & $access()$ & $next()$ & Description \\
    \cmidrule(r){2-2}\cmidrule(r){3-3}\cmidrule(r){4-4}\cmidrule(r){5-5}
    $\gamma$-code~\cite{1055349} & $2N+O(\frac{n}{h}\log N)$ & $O(h)$ &
    $O(1)$ & Sampling step $h$\\[.2em]
    $\delta$-code~\cite{1055349} &$N+2n\log\frac{N}{n}+ O(n + \frac{n}{h}\log N)$
    & $O(h)$ & $O(1)$ & Sampling step $h$\\[.2em]
    DAC~\cite{Brisaboa:2013:DBD:2400755.2401216} & $\lceil
    N(1+\frac{1}{b})\rceil+nb+o(\frac{N}{b})$ & $O(\log x_{m}/b)$ & $O(\log x_{m}/b)$ &
    parameter $b\leq \log x_{m}$\\[.2em]
    FV~\cite{FERRAGINA2007115} & $N_{\mathit{fv}}+O(\frac{n}{h}( \log N_{\mathit{fv}}+\log(h\log x_{m})))$ & $O(1)$
    & $O(1)$ & Sampling step $h$\\[.2em]
    Simple9~\cite{Anh2005}            &  $N_{\mathit{s9}}+\frac{n}{h}\log N_{\mathit{s9}}$ & $O(h)$    & $O(1)$ & Sampling step $h$\\[.2em]
    PforDelta~\cite{1617427}           &  $N_{\mathit{pfd}}+\frac{n}{h}\log N_{\mathit{pfd}}$ &  $O(h)$    & $O(1)$ & Sampling step $h$\\[.2em]
    RL & $2r(\log n-\frac{\log r}{2}+1)+o(r)$ &
    $O(\log\frac{n}{r})+O(\frac{\log^4r}{\log n})$ & $O(\frac{\log^4r}{\log n})$ &
    Number of runs, $r$ \\
    \bottomrule 	
  \end{tabular}
\end{center}
\caption{Compressed integer vectors used in the study. Given an
  integer vector $X=x_1,x_2,\ldots, x_n$, we define $x_{m}=max\{x_i | 1\leq
    i\leq n\}$, and $N=\sum_{i=1}^n(\lfloor\log x_i\rfloor+1)$ the minimum number of bits to represent all the entries of
  $X$. For Simple9, PforDelta and FV we define $N_{\mathit{s9}}$, $N_{\mathit{pfd}}$ and $N_{\mathit{fv}}=\sum_{i=1}^{n}\lfloor\log i\rfloor$
  as the number of bits needed to represent the vector $X$ with each technique
  and without sampling.}
\label{tbl:civ}
\end{table*}

For this first study of energy consumption on compact data structures we
evaluate the energy performance of one of the most basic compact data
structures, {\em Compressed Integer Vectors} (\civ). The \civ{}s are
building-blocks of more complex compact data structures, such as, 
compressed suffix arrays~\cite{SADAKANE2003294} and inverted
indexes~\cite{Anh2005}. We tested several \civ{}s present in the
literature.

Let $X = x_1, x_2, \ldots, x_n$ be the sequence of $n$ integers to encode.
A way to compress $X$ is to use statistical compression, that is, create a vocabulary of the different integers that appear in $X$, sort them by their frequency, and assign shorter codewords to those values that occur more frequently. In the case of domains where the smallest values are assumed to be more frequent, one can directly encode the numbers with a fixed instantaneous code that gives shorter codewords to smaller numbers. Thus, it is not necessary to maintain a vocabulary of symbols sorted by frequency, which may be prohibitive if the set of distinct numbers is too large. For this scenario, the best-known encoding schemes are unary codes, $\gamma$-codes, $\delta$-codes, and
Rice codes \cite{1055349,WMB99,Sol07}. 

There exist some recent fast decodable representations, such as Simple9 \cite{Anh2005}, Simple16 and PforDelta \cite{1617427}, which achieve fast decoding and little space. The approach used by these techniques is to pack a number of small integers in a computer word, using the number of bits needed by the largest number. For instance, Simple9 packs the integers of the original sequence into words of 32 bits, computing the maximum number of consecutive integers that can be included in a word using the same number of bits for all. For example it can encode 28 1-bit numbers, 14
2-bit numbers, 9 3-bit numbers, and so on. 
PForDelta can use more than 32 bits (say, 256), and treat the 10\% largest numbers as
exceptions that are encoded separately. These techniques perform very well in practice.

There exist other compression techniques based on searching for {\it runs} in the original stream of data, where a run is an interrupted sequence of the same value. When using run-length (RL) encoding, the original sequence is replaced by the representation of its runs, each of them encoded with two integers: the value that is repeated (run value) and the number of times it is repeated (run length).

Specifically, in this work we tested \civ{}s based on the $\gamma$-code and
$\delta$-code of Elias~\cite{1055349}, the {\em Directly Addressable Codes} (DAC)
of Brisaboa {\it et al.}~\cite{Brisaboa:2013:DBD:2400755.2401216}, the data structure of Ferragina and Venturini
(FV)~\cite{FERRAGINA2007115}, Simple9 of Anh and Moffat~\cite{Anh2005}, PforDelta of Zukowski
{\it et al.}~\cite{1617427} and a compressed vector based on run-length encoding
(RL). To improve the random access time of the \civ{}s
based on $\gamma$-code, $\delta$-code, Simple9 and PforDelta, we complement them with a sampling
array, storing the position of each $h$-th entry of input vector. 
The RL structure is
composed of an array $H[1..r]$ storing the first value of each run, and a
bitvector $B[1..n]$ with rank/select support, marking the beginning of each
run. With the objective of reducing the overall space of RL for vectors with long
runs, the bitvector $B$ corresponds to the sparse bitvector {\em sdarray} of
Okanohara and Sadakane~\cite{Okanohara2007}. Table~\ref{tbl:civ}
summarizes all the tested \civ{}s. The table shows the space usage of each
structure, the time complexity of the operation $\access{i}$, which recovers the
$i$-th entry on the vector, and the time complexity of the operation $\next{i}$, which recovers the $(i+1)$-th assuming that the $i$-th entry was already recovered.
The operation $\access{}$ is
preferred to random access of the entries of a \civ{}, meanwhile the operation
$\next{}$ is preferred to a left-to-right scanning of the \civ{}. In general,
the usage of the $\next{}$ operation in a sequential scanning is faster than
using the $\access{}$ operation, because we can store partial results in local
variables, since we already know the next entry to be recovered.
Note that, from the definition of the operation $\next{}$, an operation
$\access{}$ must be applied at the beginning of the left-to-right scanning. For
example, we could recover the entry at position $i+2$ using the expression
$\next{\next{\access{i}}}$.

\subsection{Access patterns}

The aim of this study case is to measure the impact in the energy consumption of
the particular data rearrangement of each \civ{} for the entries of the input
vector. For that, we performed two sets of experiments. In the first set,
we performed several binary searches over the \civ{}s. We decided to test binary search
because it is one of the most classical algorithms in Computer Science. Additionally,
binary search performs up to $\log n$ non-consecutive accesses to a vector of length
$n$, allowing us to study the impact of $\access{}$ operation. The second set of experiments
corresponds to the computation of the sum of $m$ entries of the vector. We
tested a random access pattern, choosing the $m$ entries randomly, and a
sequential access pattern, choosing the first $m$ entries in increasing
positional order. For random access we use the $\access{}$ operation and for sequential
accesses we use the $\next{}$ operation.
The sum operation is one of the most basic and efficient CPU
instructions implemented in modern computers. Thus, we argue that the resulting
values obtained in Section~\ref{subsec:results} are mainly due to the data
rearrangement of the \civ{}s instead of some overhead due to the sum operation.

\section{Experimental Evaluation}

\begin{table*}[t]
\begin{center}
  \begin{tabular}{l@{\hspace{2ex}}l@{\hspace{0.3ex}}r@{\hspace{1ex}}l@{\hspace{0.3ex}}r@{\hspace{1ex}}l@{\hspace{0.3ex}}r@{\hspace{2ex}}l@{\hspace{0.3ex}}r@{\hspace{1ex}}}
    \toprule
     & \multicolumn{2}{c}{{\tt cere}} &\multicolumn{2}{c}{{\tt kernel}} & \multicolumn{2}{c}{{\tt eins}} & \multicolumn{2}{c}{{\tt dblp}} \\
    \cmidrule(r){2-3} \cmidrule(r){4-5} \cmidrule(r){6-7} \cmidrule(r){8-9}
    \multirow{5}{*}{{\tt bwt}} & max val: &  $84$ &  max val: &  $255$ & max val: &  $252$  & max val: &  $126$\\
    & avg val: & 72 & avg val: & 77 & avg val: & 91 & avg val: & 86\\
    & max diff: &  -$78$ & max diff: &  -$223$  & max diff: &  $220$  & max diff:
    &  $111$\\
    & runs: &  $9,438,540$ &  runs: &  $2,327,498$ & runs: &  $105,892$  & runs: &  $15,060,348$\\
    \cmidrule(r){2-3} \cmidrule(r){4-5} \cmidrule(r){6-7} \cmidrule(r){8-9}
    \multirow{4}{*}{{\tt lcp}} &  max val: &  $32,469$ &  max val: &  $1,466,191$ & max val: &  $935,920$  & max val: &  $1,084$\\
    & avg val: & 2,129 & avg val: & 113,361 & avg val: & 41,322 & avg val: & 44\\
    & max diff: &  $29,562$ & max diff: &  -$1,456,752$  & max diff: &  $893,180$  & max diff: &  $961$\\
    \cmidrule(r){2-3} \cmidrule(r){4-5} \cmidrule(r){6-7} \cmidrule(r){8-9}
    \multirow{4}{*}{{\tt psi}} & max val: &  $104,857,600$ &  max val: &  $104,857,600$ & max val: &  $104,857,600$  & max val: &  $104,857,600$\\
    & avg val: & 52,428,800 & avg val: & 52,428,800 & avg val: & 52,428,800 & avg val: & 52,428,800 \\
    & max diff: &  -$104,857,295$ & max diff: &  -$104,851,818$  & max diff: &  -$104,846,882$  & max diff: &  -$104,854,855$\\
    \bottomrule 	
  \end{tabular}
\end{center}
\caption{Statistics of the datasets.}
\label{tbl:datasets}
\end{table*}

\subsection{Experimental Setup}
In our experiments, we set the sampling steps of the tested \civ{}s to values
that offer a good trade-off between space and access time. Thus, for the \civ{}s
$\delta$-code, $\gamma$-code, FV ({\tt fv}) and Simple9 ({\tt s9}), the sampling
value was 128, and for PforDelta ({\tt pfd}) the sampling value was 1024. The
parameter $b$ of DAC ({\tt dac}) was chosen by an optimization
algorithm~\cite{Brisaboa:2013:DBD:2400755.2401216} that computes a different $b$ value for each level of the representation, reducing the
final size of the representation.

Depending on the properties of the integer sequence, it is more convenient sometimes to store the differences between two consecutive elements of the vector instead of the elements themselves. To test such situation, we
implemented modified versions of the \civ{}s of Table~\ref{tbl:civ} to store the
differences. To deal with negative differences we use the {\em ZigZag encoding}\footnote{{\em ZigZag encoding} maps signed integers to unsigned integers using a ``zig-zag'' strategy over the positive and negative integers, so that numbers with small absolute values are mapped to small unsigned integers. For example, -1 is encoded as 1, 1 is encoded as 2, -2 is encoded as 3, 2 is encoded as 4, and so on.}.
For the rest of the document, we identify the modified versions with the suffix
{\tt \_zz}.

To measure the energy efficiency of the \civ{}s, we tested two datasets. The first group corresponds to a sorted integer vector of length 104,857,600, called {\tt sorted}.
The elements of the vector were
randomly selected from the range $[0.. 2^{30}]$. This dataset will be used to test the binary search algorithm. The second group corresponds to four datasets from the {\em Pizza\&Chili
  Corpus}: {\it i)} {\tt dblp}, an XML document of publications on Computer Science; {\it ii)}
the repetitive datasets {\tt cere}, the DNA sequencing of the yeast
Saccharomyces Cerevisiae; {\it iii)} {\tt kernel}, a collection of Linux Kernels; and {\it iv)} {\tt
  eins}, the English versions up to November of 2006 of the Wikipedia article of
Albert Einstein. For each dataset, we extracted the first 100 MB of data,
equivalent to 104,857,600 symbols. In order to study different types of vectors,
we computed the Burrows-Wheeler Transform ({\tt bwt}), the longest common prefix
array ({\tt lcp}) and the $\Psi$ function used in compressed suffix arrays
({\tt psi}). The {\tt bwt} vectors have ranges of equal letters, called runs,
especially for repetitive datasets; for non-repetitive datasets, such as {\tt
  dblp}, most of the entries of the {\tt lcp} vectors are small compared to the length of the vector;
  and the {\tt psi} vectors have ranges of
increasing values, especially for repetitive datasets. This second group of datasets will
be used to test the random and sequential access patterns of the sum of entries. Table~\ref{tbl:datasets}
shows some statistics of the datasets and Table~\ref{tbl:space} shows the space
consumption of the \civ{}s for all the datasets. The results correspond to the expected behaviour of each \civ{} taking into account the distribution of values of each dataset. {\tt rl} obtains the best compression for those {\tt bwt} datasets with a low number of runs, that is, for those datasets with longer runs, namely {\tt kernel} and {\tt eins}. For {\tt lcp} datasets, the techniques that obtain the best results are {\tt dac} and {\tt dac\_zz}. {\tt pfd\_zz} is the method achieving the smallest representations for the rest of the datasets, and also the one that obtains the best overall performance. 

\begin{table*}[t]
\begin{center}
  \begin{tabular}{l@{\hspace{2ex}}r@{\hspace{2ex}}r@{\hspace{2ex}}r@{\hspace{2ex}}r@{\hspace{2ex}}r@{\hspace{2ex}}r@{\hspace{2ex}}r@{\hspace{2ex}}r@{\hspace{2ex}}r@{\hspace{2ex}}r@{\hspace{2ex}}r@{\hspace{2ex}}r@{\hspace{2ex}}r@{\hspace{2ex}}}
    \toprule
    & \multicolumn{3}{c}{{\tt cere}} & \multicolumn{3}{c}{{\tt kernel}} &
    \multicolumn{3}{c}{{\tt eins}} & \multicolumn{3}{c}{{\tt dblp}} \\
    \cmidrule(r){2-4} \cmidrule(r){5-7} \cmidrule(r){8-10}\cmidrule(r){11-13}
    & {\tt bwt} & {\tt psi} & {\tt lcp} & {\tt bwt} & {\tt psi} & {\tt lcp} & {\tt bwt} &
             {\tt psi} & {\tt lcp} &  {\tt bwt} & {\tt psi} & {\tt lcp} & {\tt sorted}\\
    \midrule
    {\tt plain} & 100.0 & 400.0 & 400.0  & 100.0  & 400.0 & 400.0 & 100.0 & 400.0 & 400.0 & 100.0 & 400.0 & 400.0 & 400.0\\
    \midrule
    {\tt $\delta$-code} &  140.5 & 424.6  & 194.5 & 134.1 & 424.6  & 268.4  & 137.9 &  424.6  & 276.4 & 135.7 & 424.6 & 122.2 & 440.6\\
    {\tt $\gamma$-code} & 165.5 & 633.4 & 233.9 & 152.7 & 633.4 & 352.1 & 160.3 & 633.4 & 364.9 & 155.9 & 633.4 & 129.0 & 665.5\\
    {\tt dac}           & 87.5 & 337.5 & \underline{160.0} & 100.0 & 337.5 & \underline{228.0} & 100.0 & 337.5 & 213.8 & 87.5 & 337.5 & 94.0 & 350.0\\
    {\tt fv}            & 203.1 & 462.1 & 249.8 & 209.2 & 462.1 & 321.4 & 200.5 & 462.1 & 327.7 & 198.3 & 462.1 & 197.4 & 478.1\\
    {\tt s9}            &  102.9 & 403.1 & 193.4 & 100.0 & 403.1 & 349.2 & 102.7 & 403.1 & 352.7 & 101.8 & 403.1 & 94.9 & 403.1\\
    {\tt rl}            & 44.0 & 439.1 & 446.2 & \underline{11.4} & 439.1 & 450.2 & \underline{0.6} & 439.1 & 440.8 & 68.6 & 439.1 & 309.6 & 439.1\\
    {\tt pfd}           & 88.2 & 399.2 & 164.1 & 84.2 & 399.1 & 234.5 & 85.8 & 399.1 & 213.6 & 83.8 & 399.2 & 78.8 & 400.1\\
    \midrule
    {\tt $\delta$-code\_zz} & 26.8 & 63.0  & 204.5 & 20.6 & 58.6  & 303.9 & 18.4 & 56.1  & 249.3 & 31.3 & 64.1 & 66.5 & 72.6\\
    {\tt $\gamma$-code\_zz} & 27.0 & 52.6 & 251.7 & 24.2 & 47.0 & 406.6 & 25.0 & 43.7 & 321.0 & 31.3 & 54.0 & 94.0 & 65.1\\

{\tt dac\_zz}           &  35.3 & 47.8 & 169.4 & 30.5 & 43.8 & 250.0 & 28.8 & 41.4 & \underline{213.1} & 39.1 & 50.3 & \underline{65.2} & 59.7\\
    {\tt fv\_zz}            & 532.4 & 570.2 & 657.2 & 541.8 & 567.4 & 747.1 &
    540.7 & 565.7 & 704.2  & 534.6 & 570.9 & 563.5 & 551.5\\
    {\tt s9\_zz}            & 62.1 & 72.6  & 203.0 & 31.9 & 44.6 & 382.5 & 20.5 & 34.9 & 324.5 & 37.6 & 49.0 & 78.7 & 65.8\\
    {\tt pfd\_zz}           & \underline{25.9} & \underline{38.9} & 169.1 & 17.4 & \underline{31.1} & 253.8 & 13.8 & \underline{26.5} & 216.0 & \underline{27.2} & \underline{39.2} & 65.5 & \underline{54.8}\\
    \bottomrule 	
  \end{tabular}
\end{center}
\caption{Space usage of the compressed vectors for the datasets of
  Table~\ref{tbl:datasets}, in megabytes. Smallest values are
  underlined.}
\label{tbl:space}
\end{table*}

\begin{figure*}[t]
  \centering
  \begin{subfigure}[b]{0.33\textwidth}
    \includegraphics[width=0.985\linewidth]{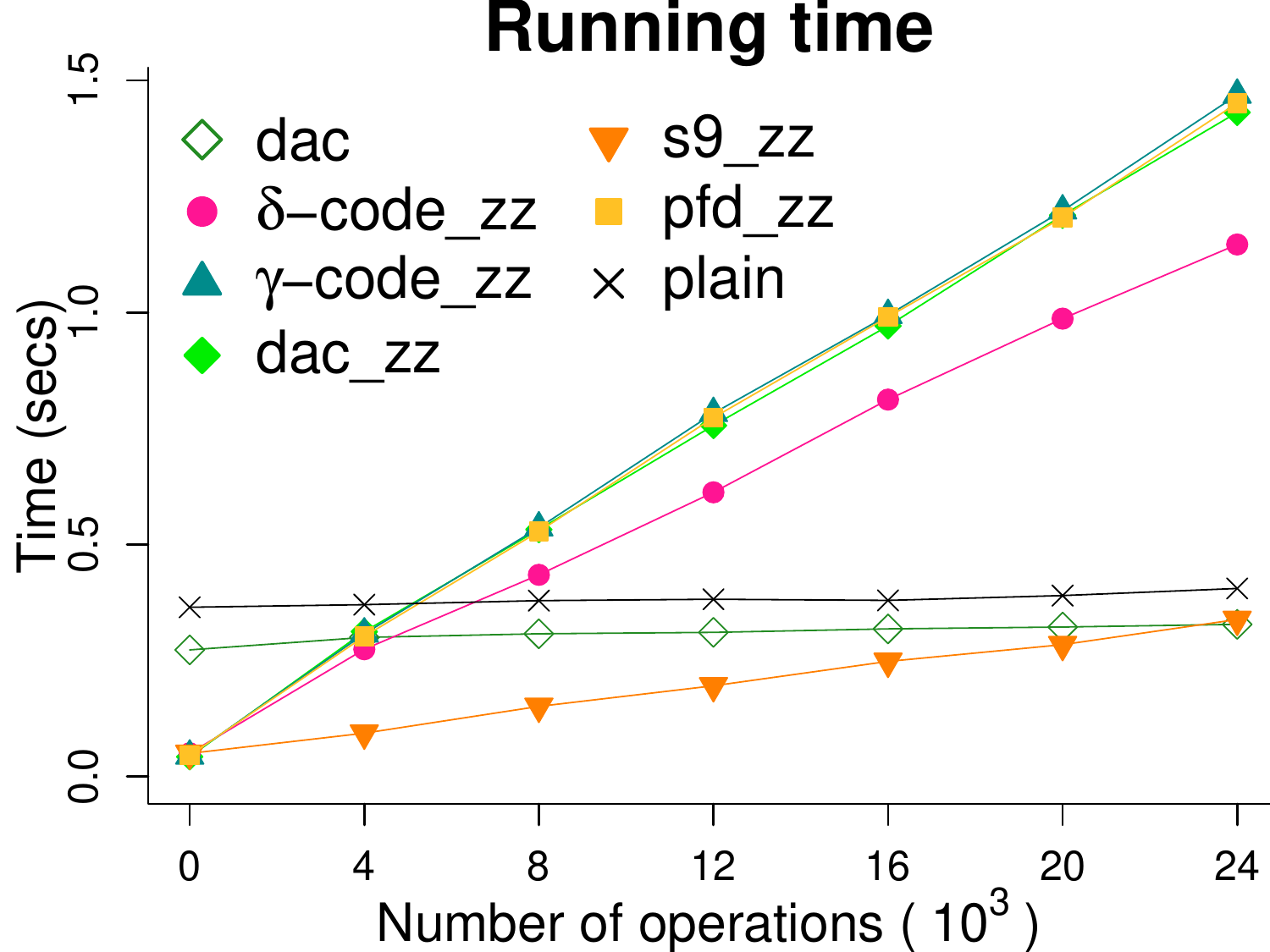}
  \end{subfigure}\hspace{2em}%
  \begin{subfigure}[b]{0.33\textwidth}
    \includegraphics[width=0.985\linewidth]{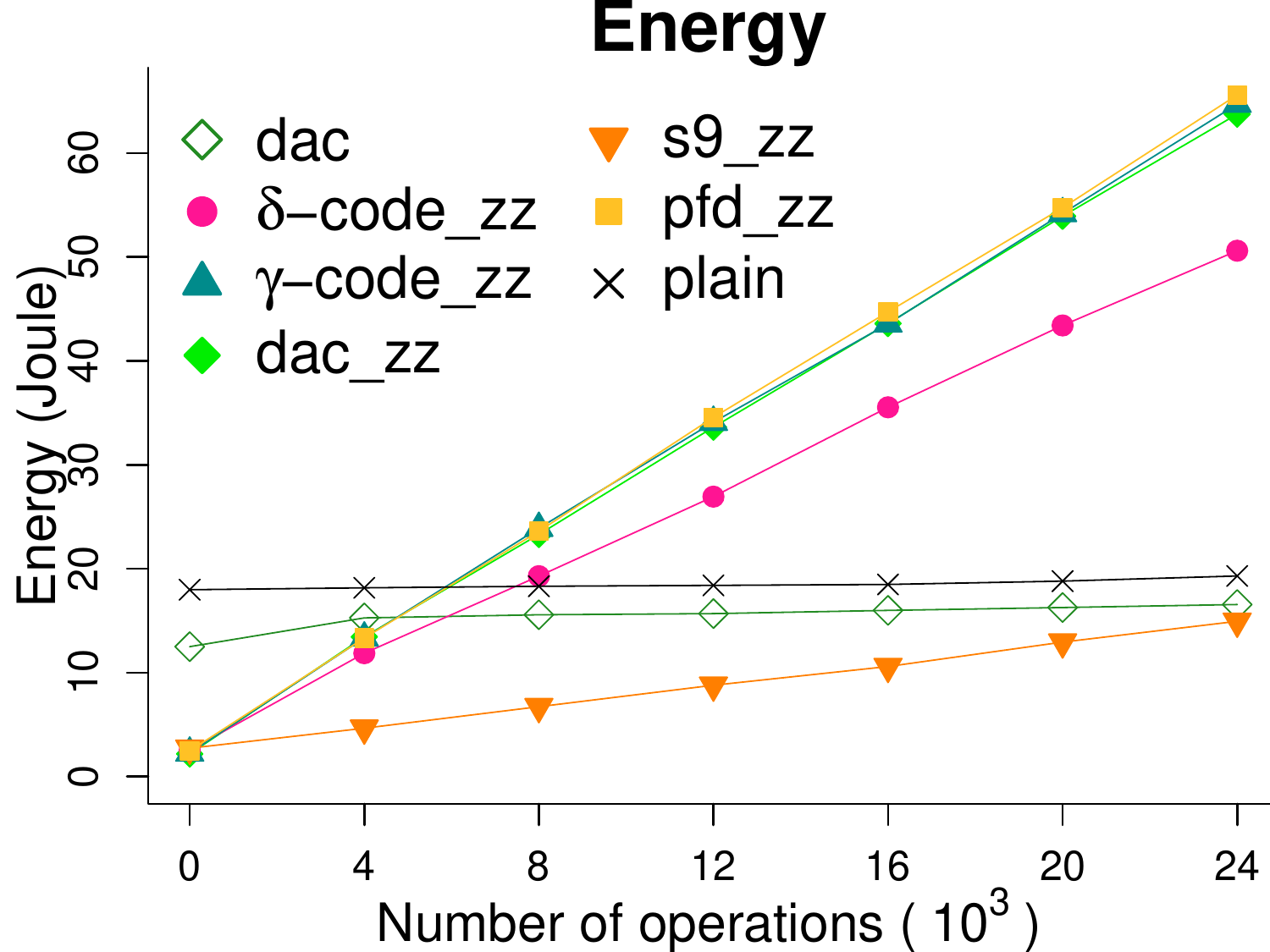}
  \end{subfigure}
  \caption{Running time and energy consumption of binary search}
  \label{fig:binsearch}
\end{figure*}

The experiments were carried out on a Non-uniform memory access (NUMA) machine with two NUMA nodes or packages. Each
NUMA node includes a 8-core Intel Xeon CPU
(E5-2470) processor clocked at 2.3 GHz. The machine runs Linux
4.9.0-6-amd64, in 64-bit mode. Each core has L1i, L1d and L2
caches of size 32 KB, 32 KB and 256 KB, respectively. All the cores of a NUMA node
share a L3 cache of 20 MB. With respect to the associativity of each cache, which
may impact in the energy consumption, L1i, L1d and L2 are 8-way
set-associative, and L3 is 20-way set-associative. All cache levels implement
the write-back policy.
Each NUMA node has a 31 GB DDR3 RAM memory, clocked at
1067 MHz. Hyperthreading was enabled, giving a total of 16 logical cores per NUMA
node. The algorithms were implemented in C++ and compiled with GCC and -O3
optimization flag. According to the work of Kambadur and
Kim~\cite{Kambadur:2014:ESE:2660193.2660196}, the optimization flag -O3 of GCC
offers significant energy savings.
We measured the running time using the {\tt clock\_gettime}
function at nanosecond resolution. Each experiment was repeated 10 times and the
median is reported.

\subsection{Power Measurement}
To measure the energy consumption of the \civ{}s, we used the {\em Intel
  RAPL (Running Average Power Limit)} Interface~\cite{intelDoc}. The Intel RAPL
Interface aggregates the content of several 
specialized registers, called {\em Model Specific Registers (MSR)}, to provide an
estimation of the energy consumption at cores level (all cores in a processor),
package level (all cores in a processor, memory controller, last level cache, among
other components) and DRAM memory level. Depending on the processor
model of the machine executing the experiments, we may have access to
the energy estimation of only the cores level and package level, which is our
case. In this work we report the energy estimation of the package level.\footnote{We replicated part of our experimental in a machine with reduced memory capacity and with access to the three levels of energy estimations (cores, package and DRAM memory). We observed that, on average, the energy consumption of DRAM memory corresponds to 5\% for sequential access pattern and to 7\% for random access pattern of the total energy consumption. Therefore, for this work, an analysis considering only the package level is valid.}
The Intel RAPL Interface has been used 
in previous energy-consumption studies and it has provided reasonably accurate
measurements \cite{Hackenberg2013,Khan:2018:RAE:3199681.3177754}.
We used the Linux profiler {\em Perf} to report the energy estimations of Intel
RAPL. Perf also allows us to measure more
metrics, such as, CPU cycles, cache hits and cache misses of the L1 cache and
the last level cache, number of instructions, etc. We used Perf 4.9.110 in the
experiments.

\subsection{Results and Discussion}
\label{subsec:results}

\begin{figure*}[t]
  \centering
  \begin{subfigure}[b]{0.33\textwidth}
    \includegraphics[width=0.985\linewidth]{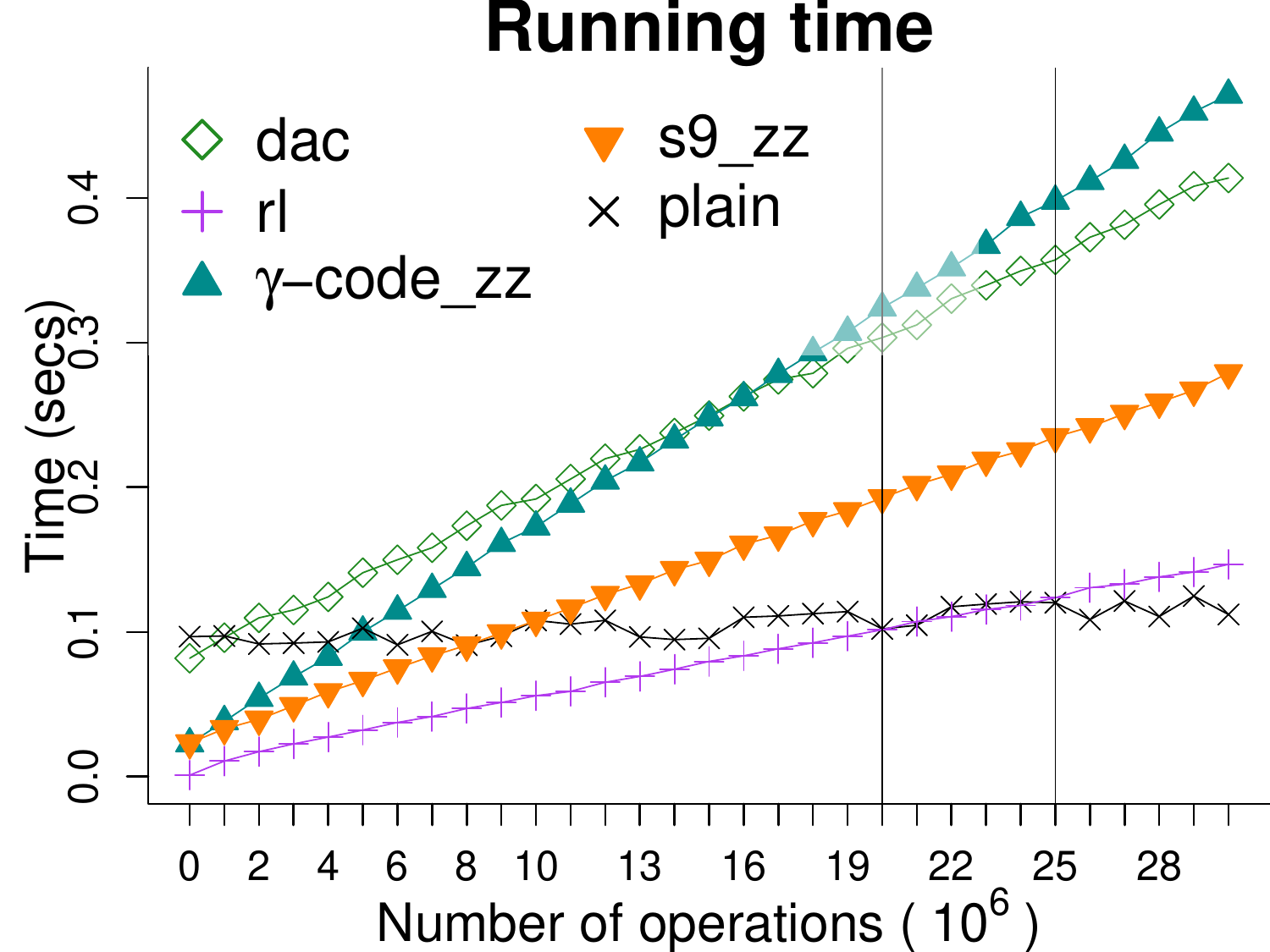}
  \end{subfigure}\hspace{0em}%
  \begin{subfigure}[b]{0.33\textwidth}
    \includegraphics[width=0.985\linewidth]{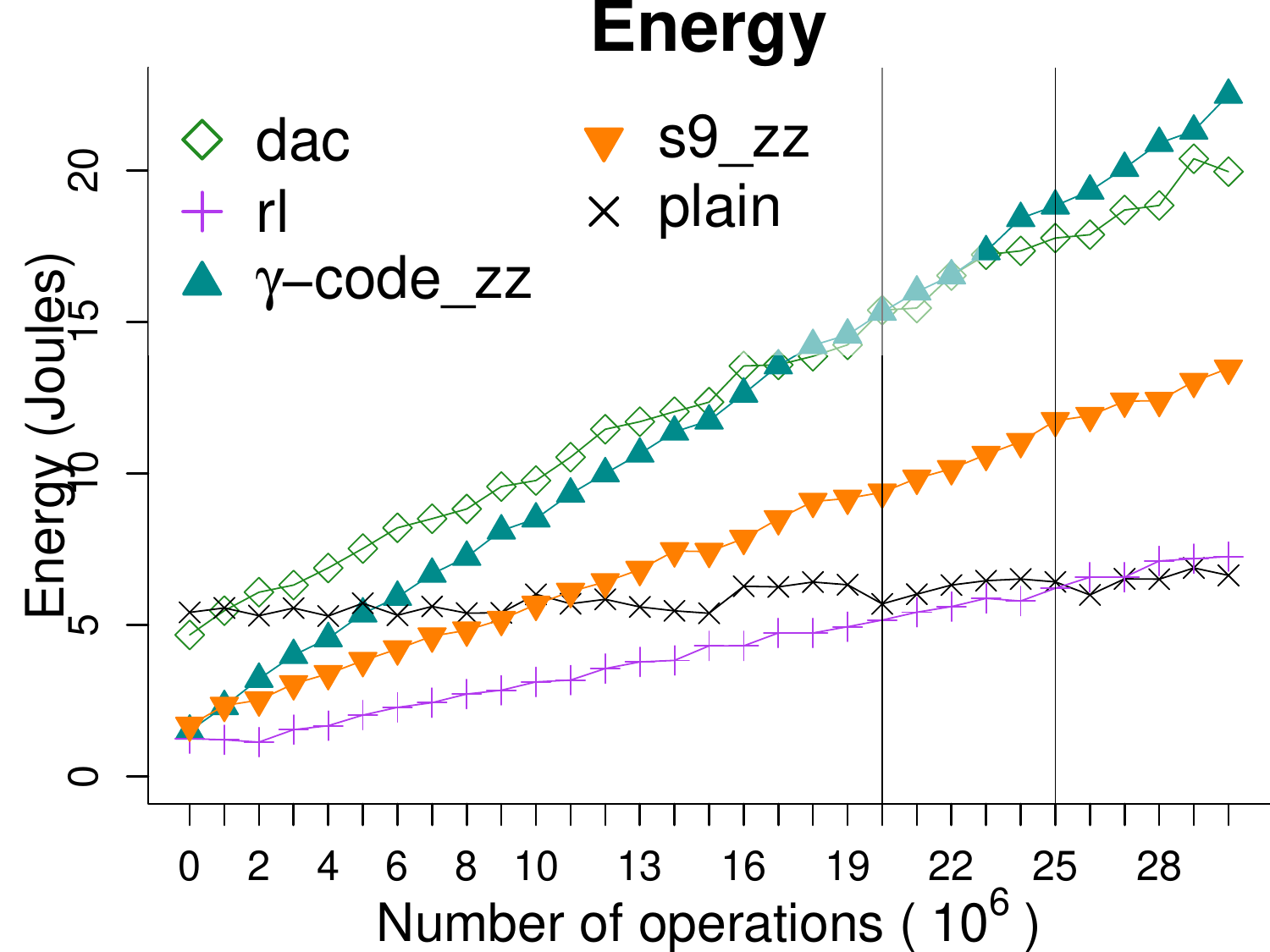}
  \end{subfigure}\hspace{0em}%
  \begin{subfigure}[b]{0.33\textwidth}
    \includegraphics[width=0.985\linewidth]{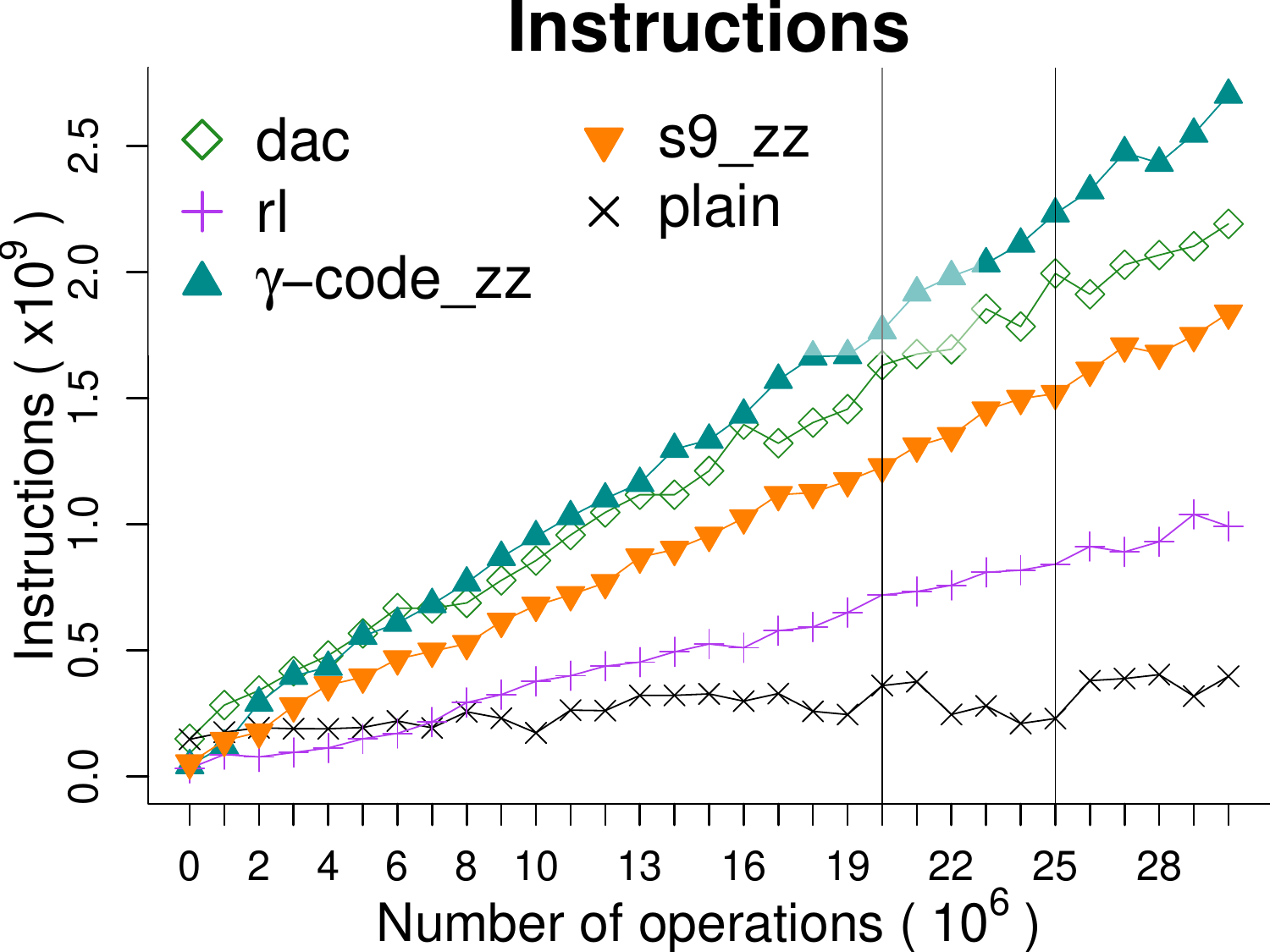}
  \end{subfigure}
  \begin{subfigure}[b]{0.32\textwidth}
    \includegraphics[width=0.985\linewidth]{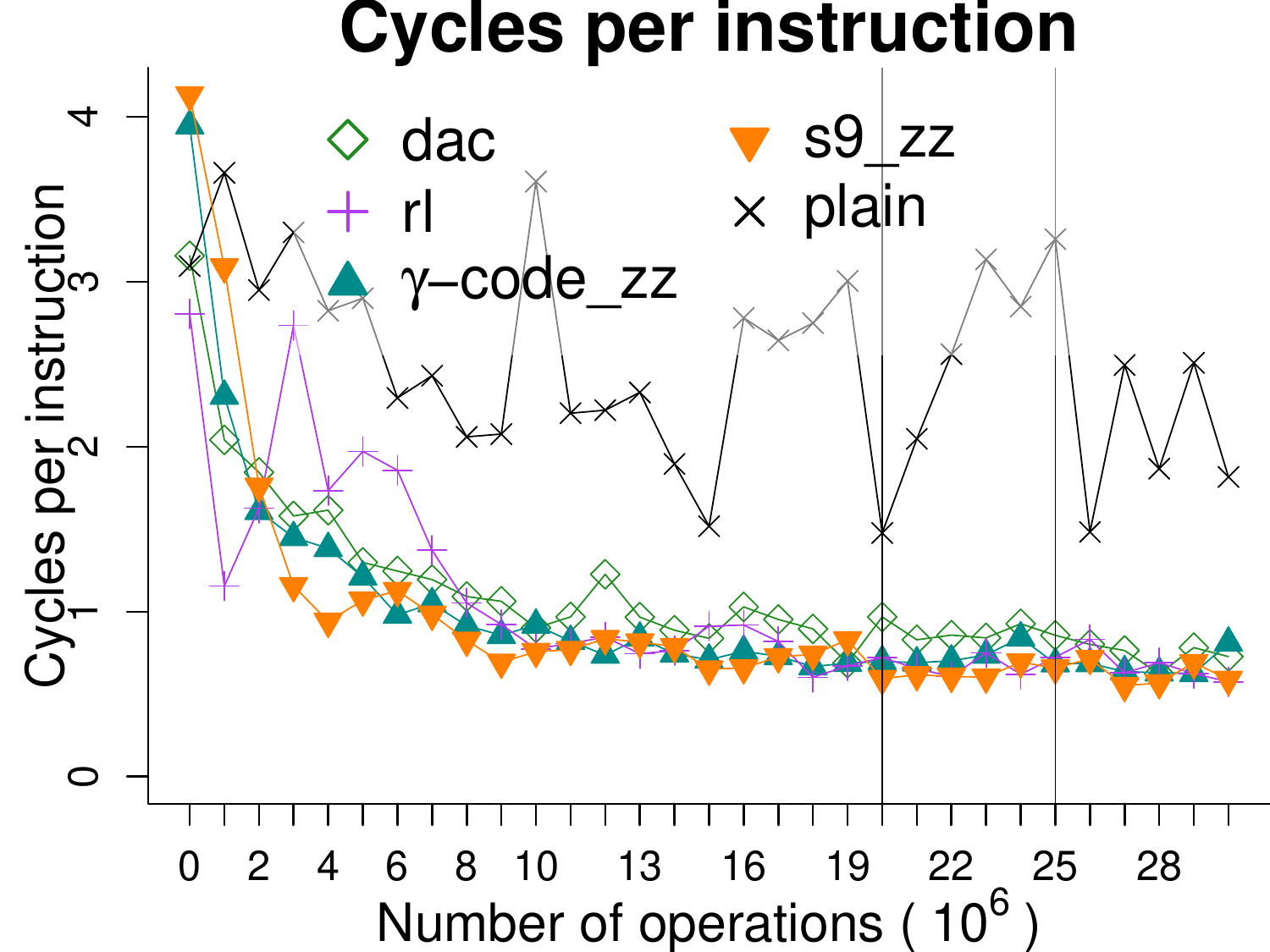}
  \end{subfigure}
  \begin{subfigure}[b]{0.32\textwidth}
    \includegraphics[width=0.985\linewidth]{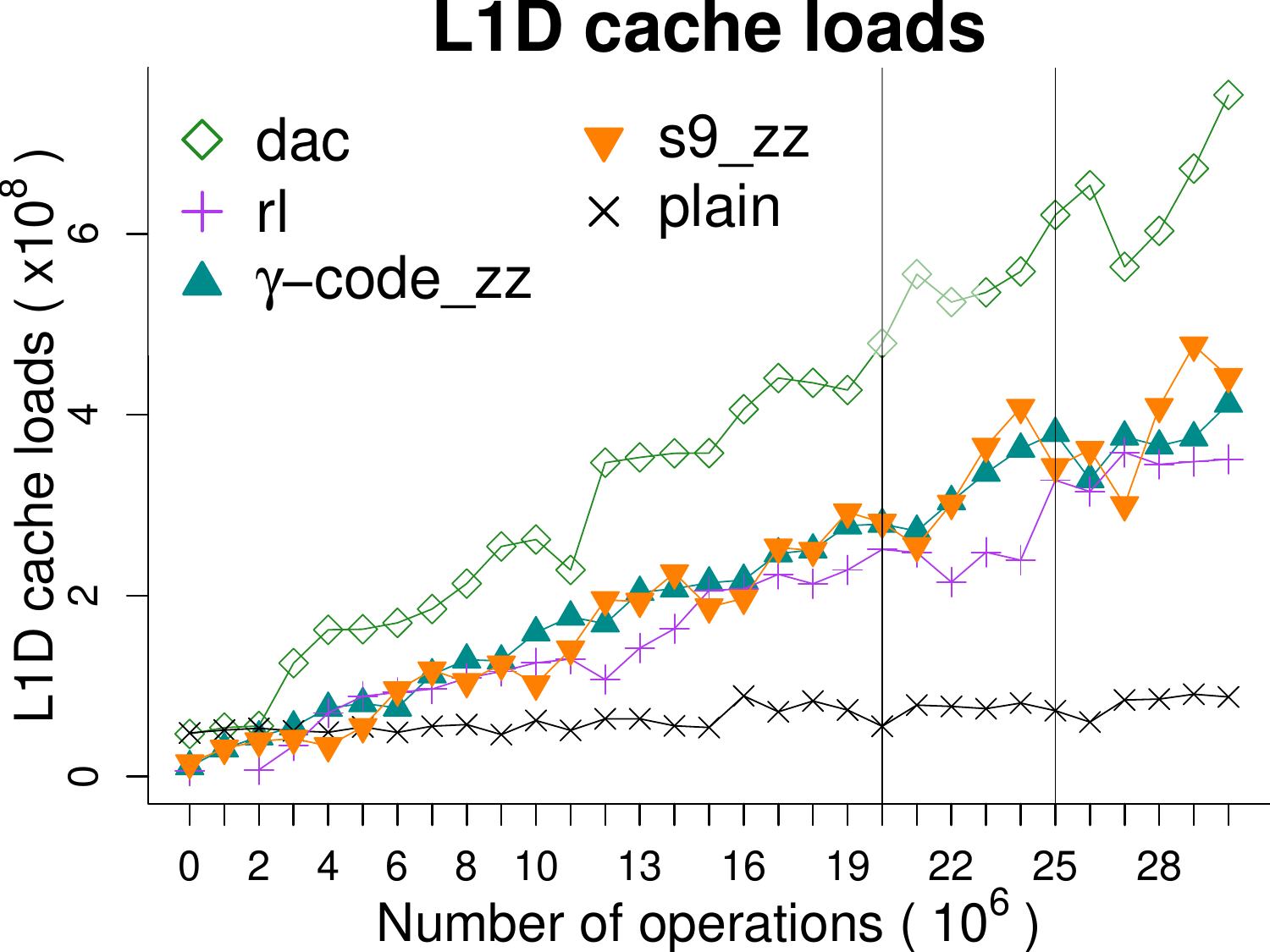}
  \end{subfigure}
  \begin{subfigure}[b]{0.32\textwidth}
    \includegraphics[width=0.985\linewidth]{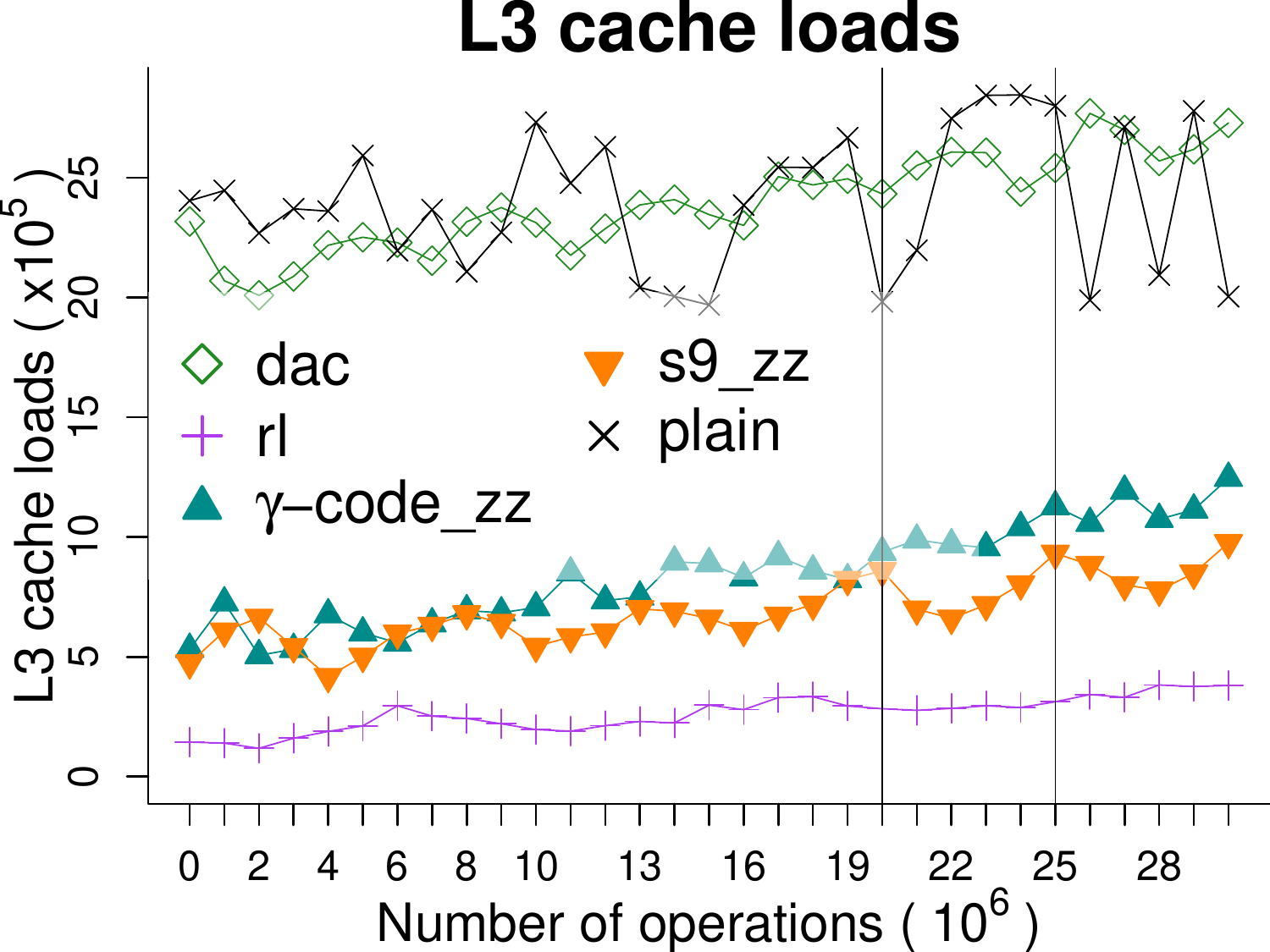}
  \end{subfigure}
  \begin{subfigure}[b]{\textwidth}
    \caption{{\tt bwt} vector of the dataset {\tt eins}}
    \label{fig:seq-bwt-eins}
  \end{subfigure}
  \begin{subfigure}[b]{0.33\textwidth}
    \includegraphics[width=0.985\linewidth]{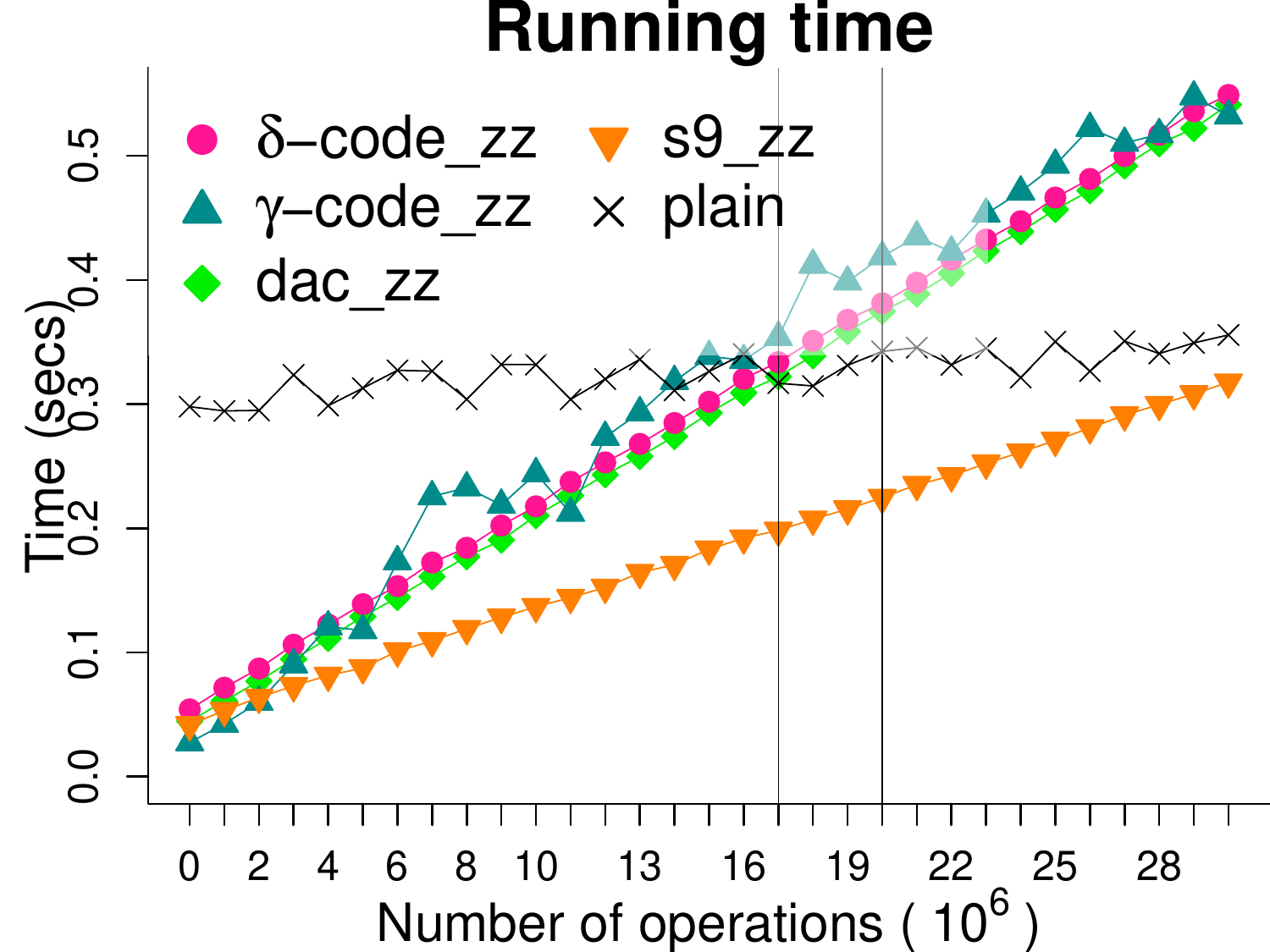}
  \end{subfigure}\hspace{0em}%
  \begin{subfigure}[b]{0.33\textwidth}
    \includegraphics[width=0.985\linewidth]{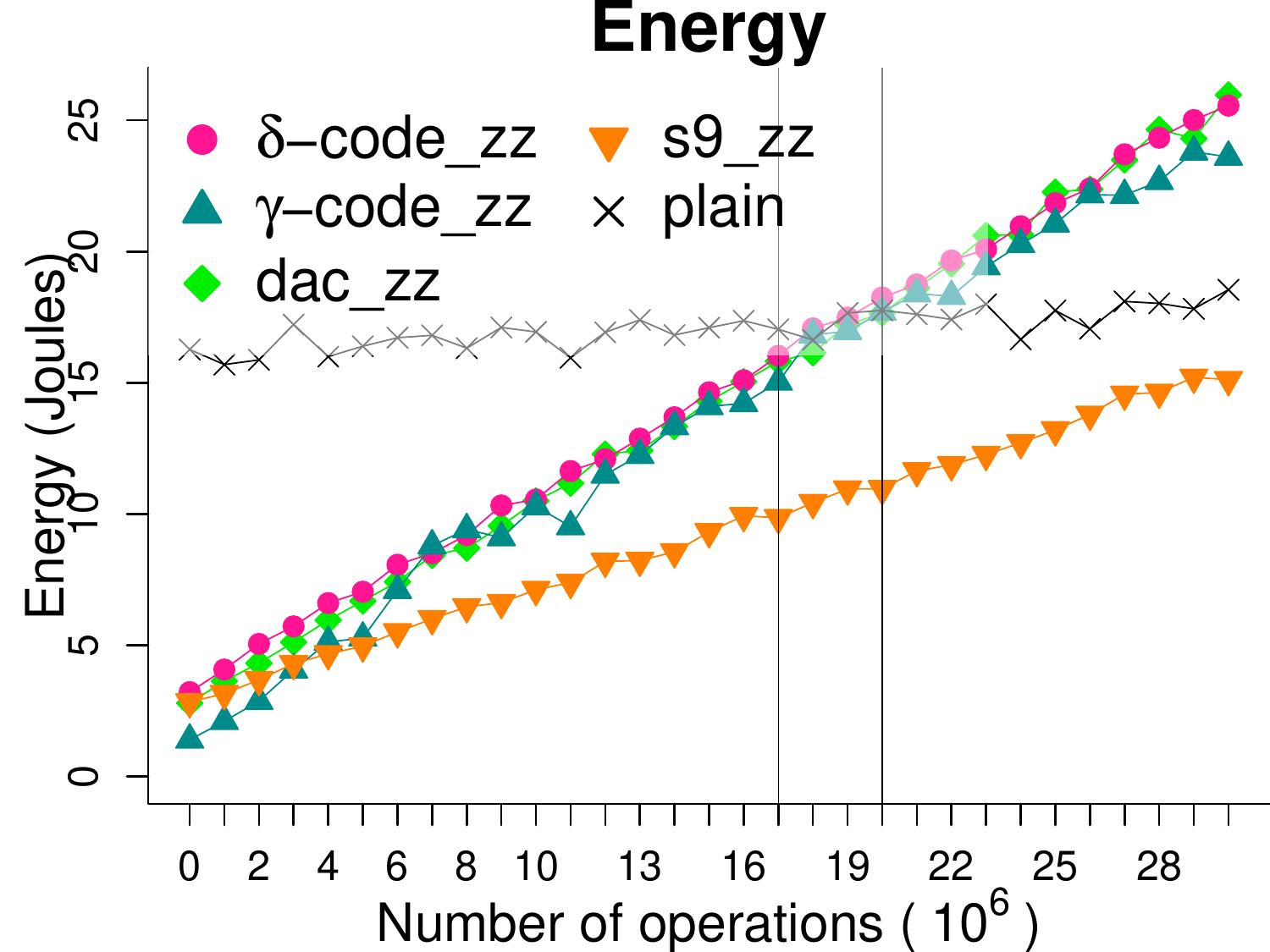}
  \end{subfigure}\hspace{0em}%
  \begin{subfigure}[b]{0.33\textwidth}
    \includegraphics[width=0.985\linewidth]{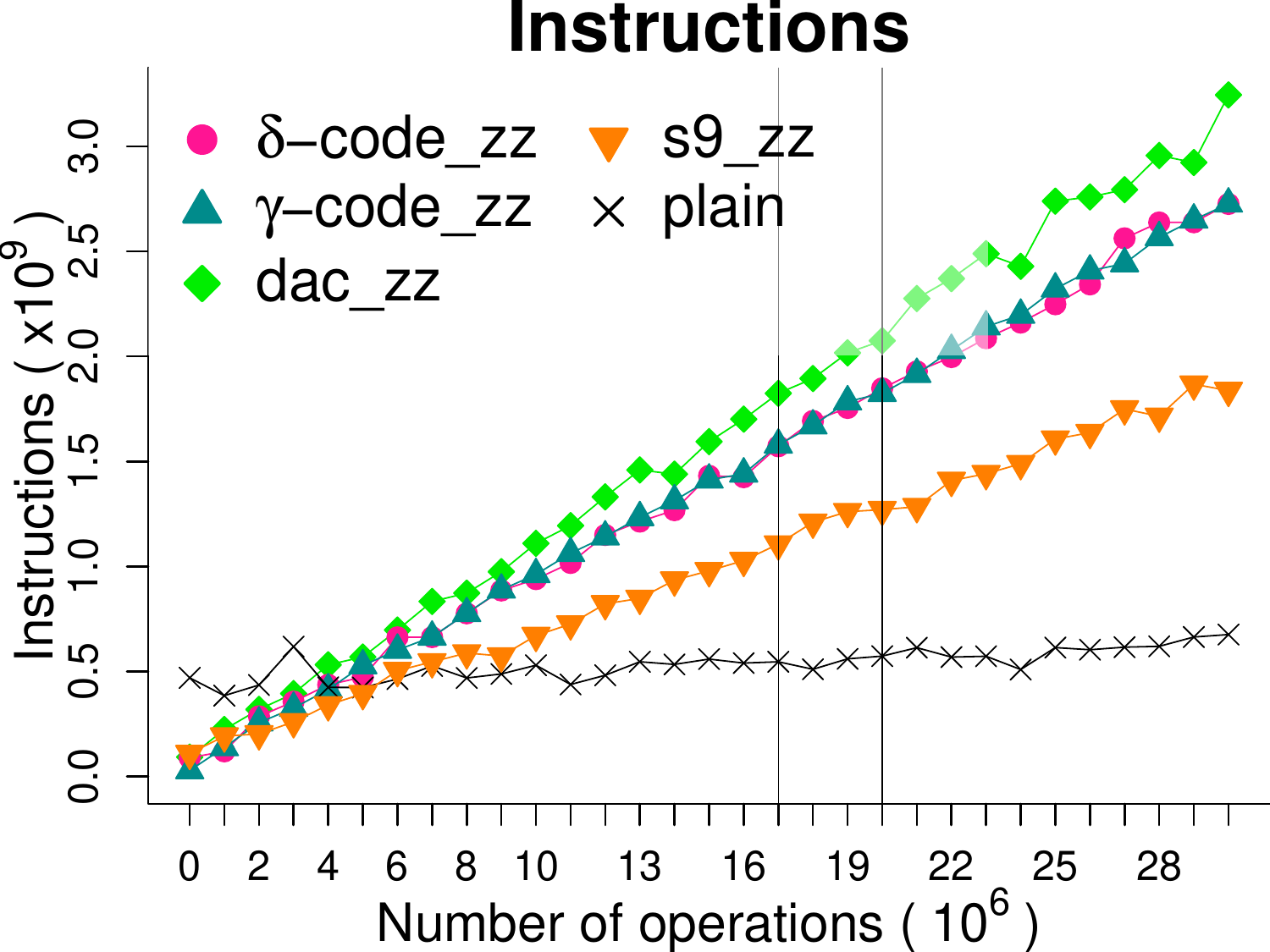}
  \end{subfigure}
  \begin{subfigure}[b]{0.32\textwidth}
    \includegraphics[width=0.985\linewidth]{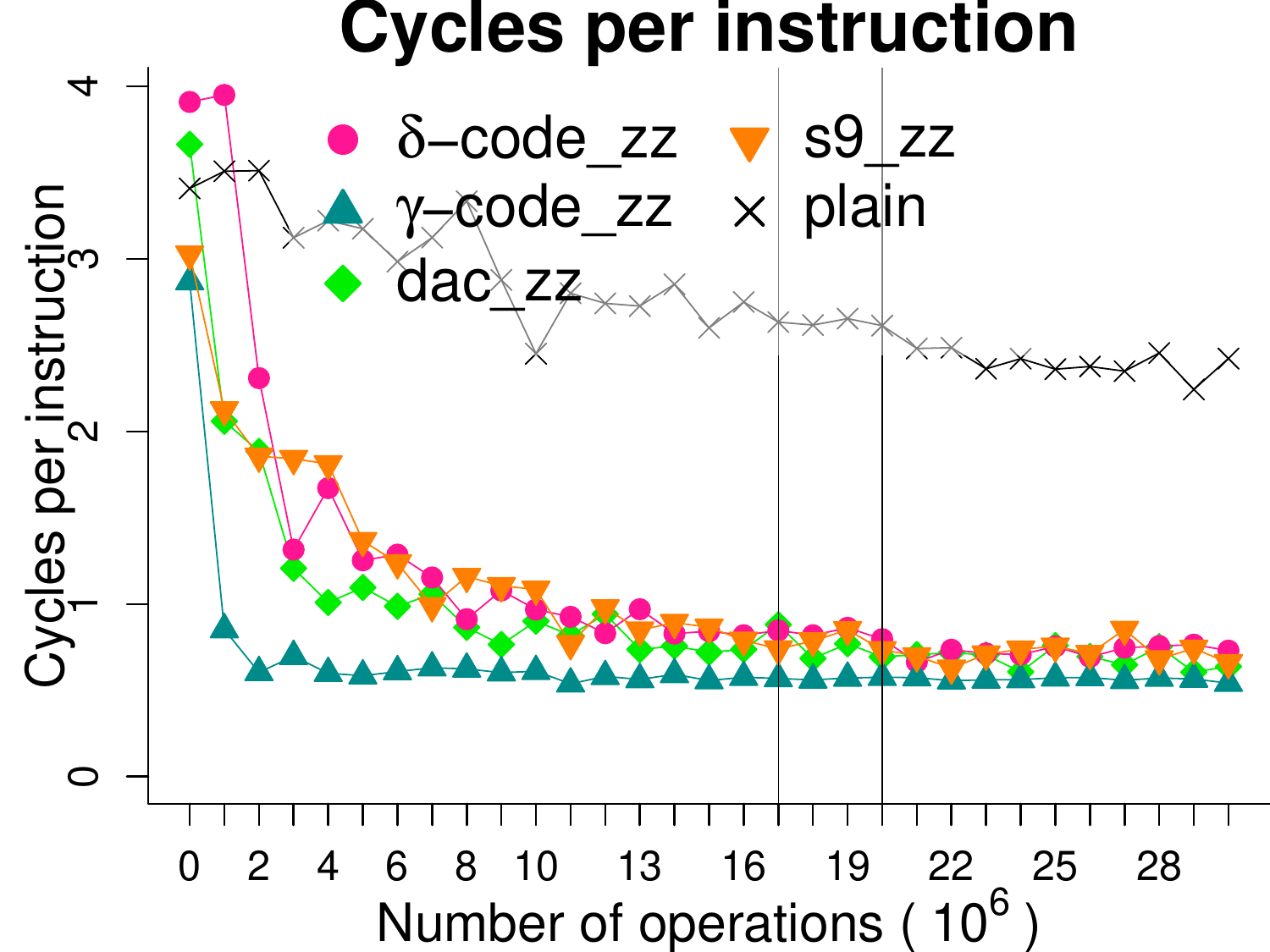}
  \end{subfigure}
  \begin{subfigure}[b]{0.32\textwidth}
    \includegraphics[width=0.985\linewidth]{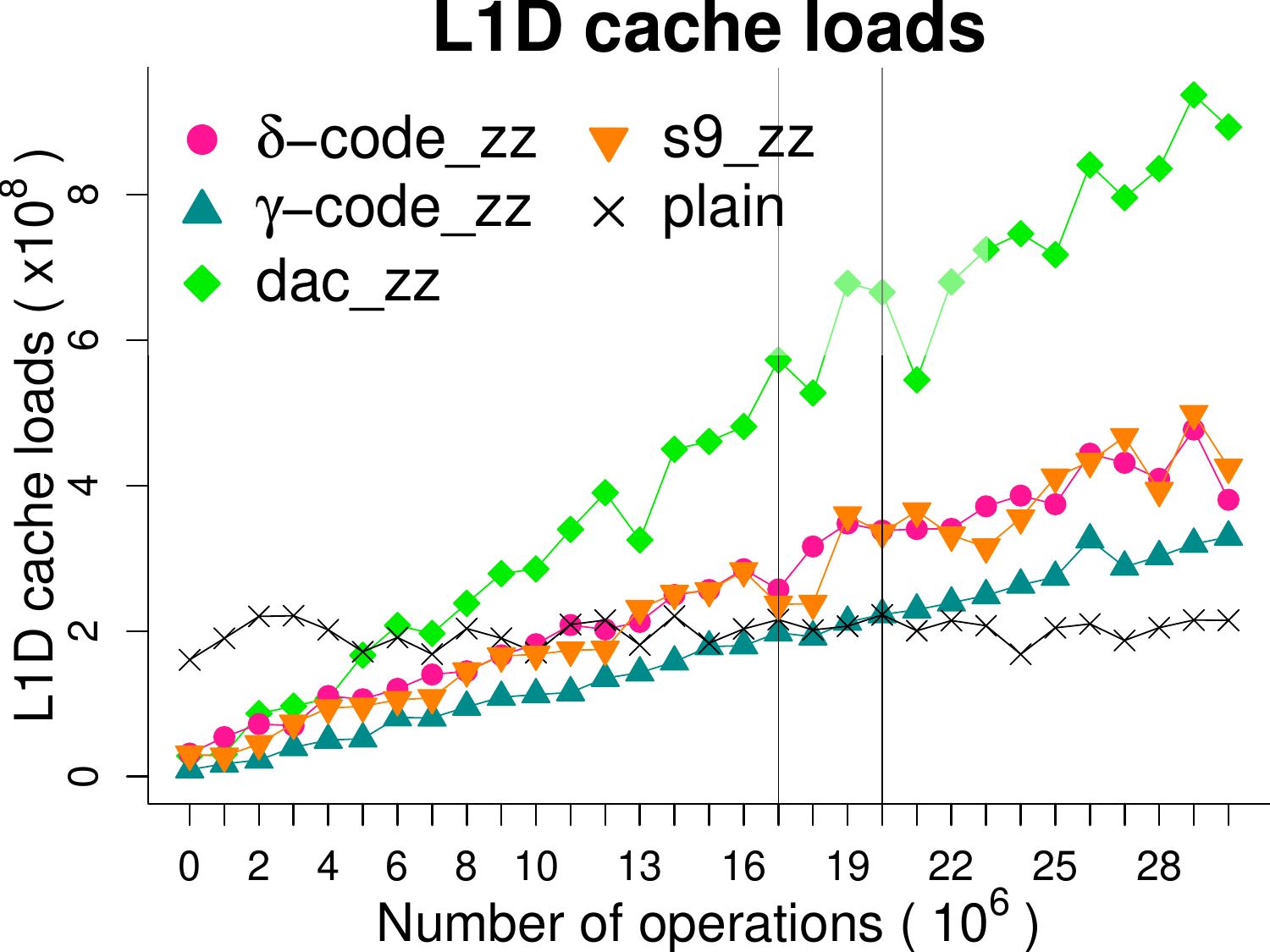}
  \end{subfigure}
  \begin{subfigure}[b]{0.32\textwidth}
    \includegraphics[width=0.985\linewidth]{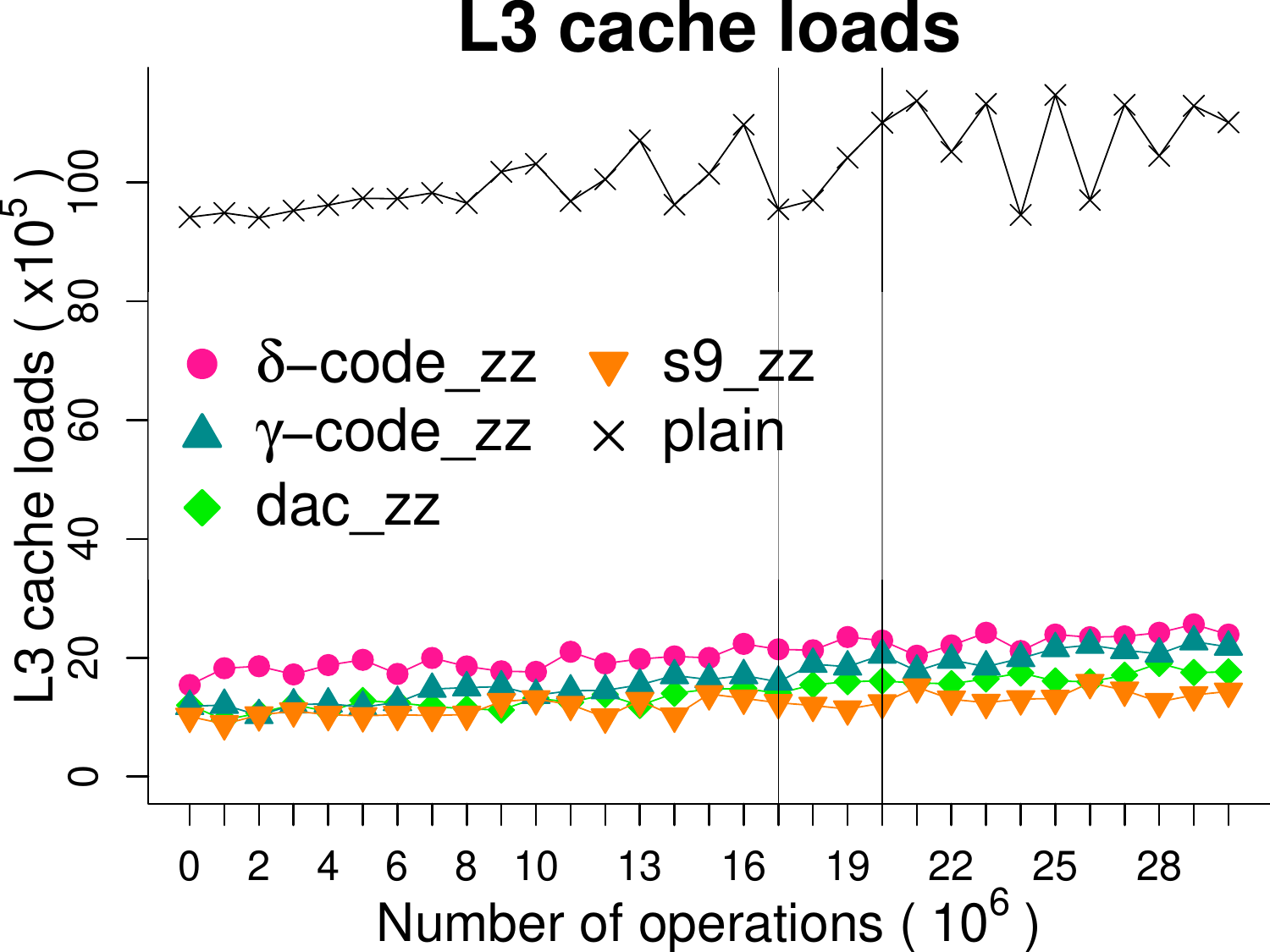}
  \end{subfigure}
  \begin{subfigure}[b]{\textwidth}
    \caption{{\tt psi} vector of the dataset {\tt kernel}}
    \label{fig:seq-kernel-psi}
  \end{subfigure}
  
  \caption{Experiments with sequential access pattern (Part I)}
  \label{fig:sequential_access_1}
\end{figure*}

\begin{figure*}[t]
  \centering
  \begin{subfigure}[b]{0.33\textwidth}
    \includegraphics[width=0.985\linewidth]{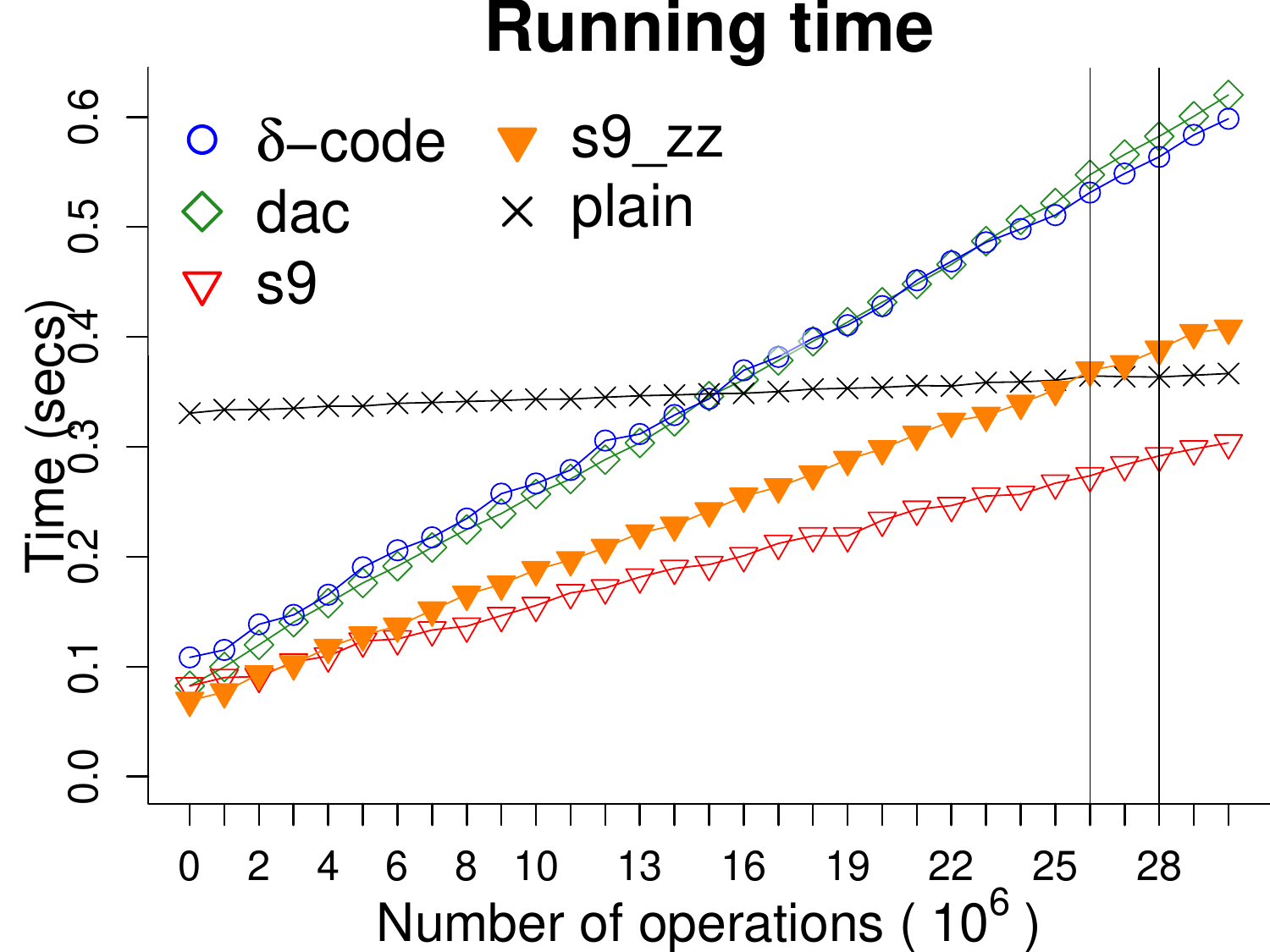}
  \end{subfigure}\hspace{0em}%
  \begin{subfigure}[b]{0.33\textwidth}
    \includegraphics[width=0.985\linewidth]{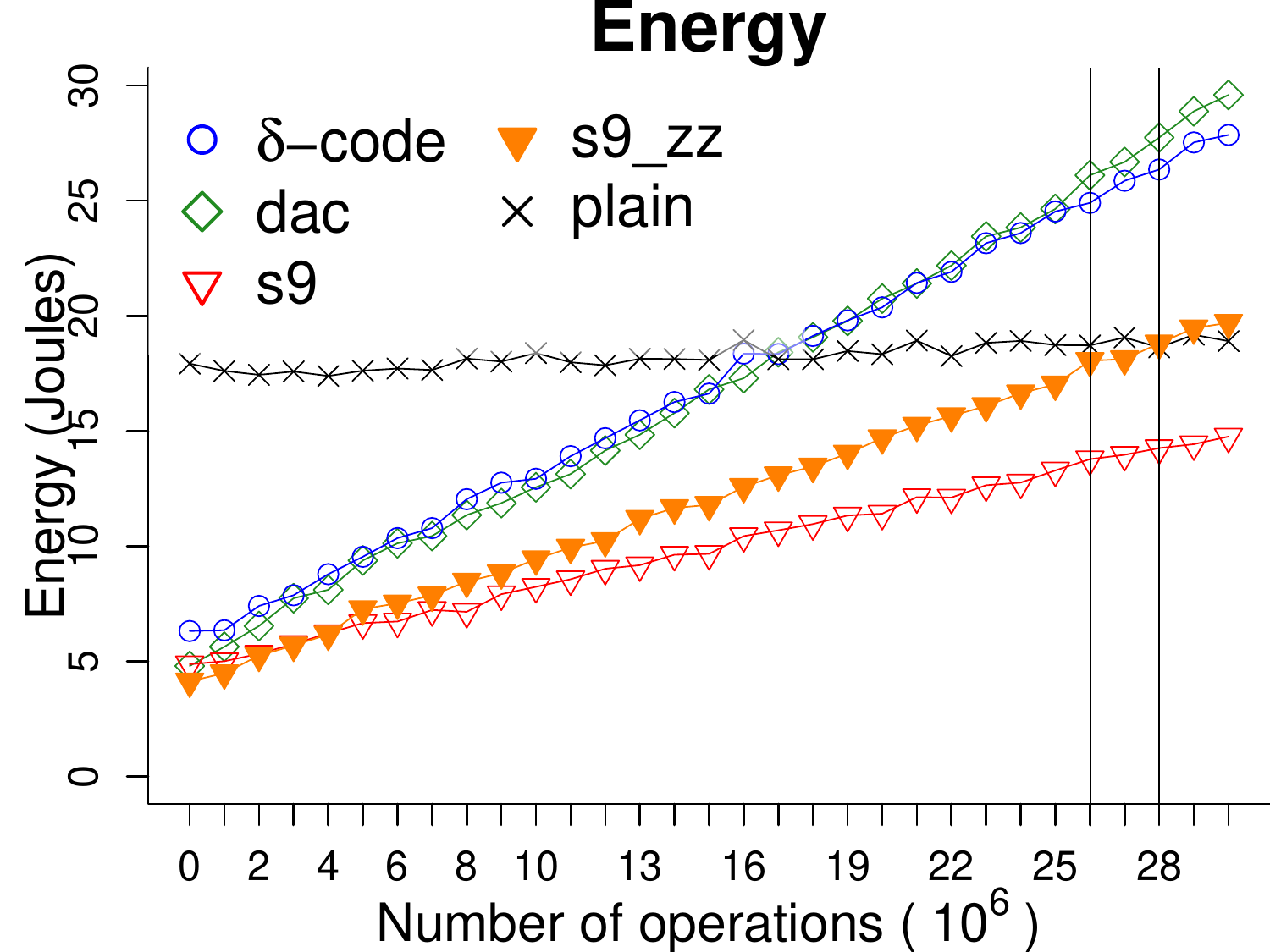}
  \end{subfigure}\hspace{0em}%
  \begin{subfigure}[b]{0.33\textwidth}
    \includegraphics[width=0.985\linewidth]{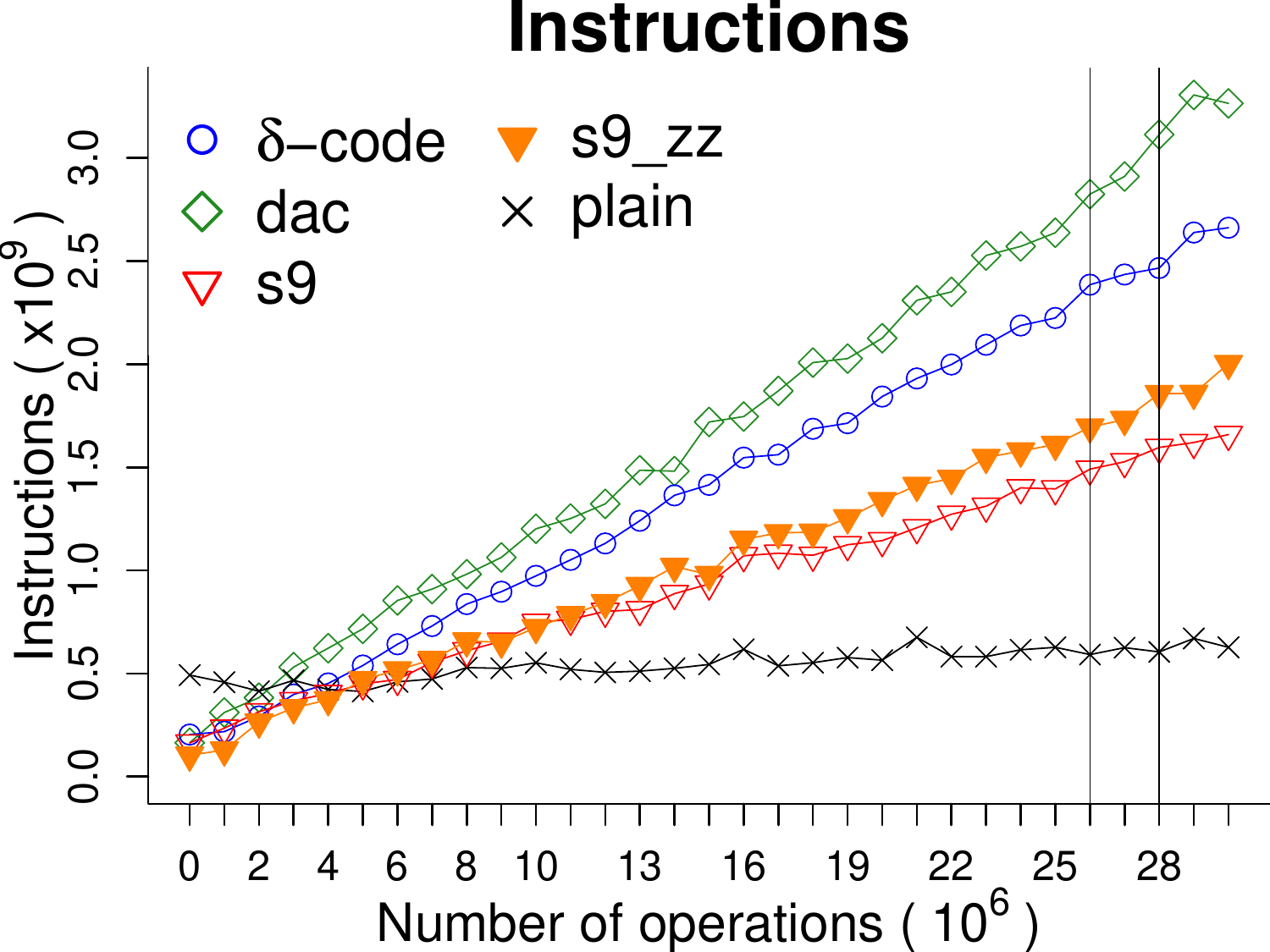}
  \end{subfigure}
  \begin{subfigure}[b]{0.32\textwidth}
    \includegraphics[width=0.985\linewidth]{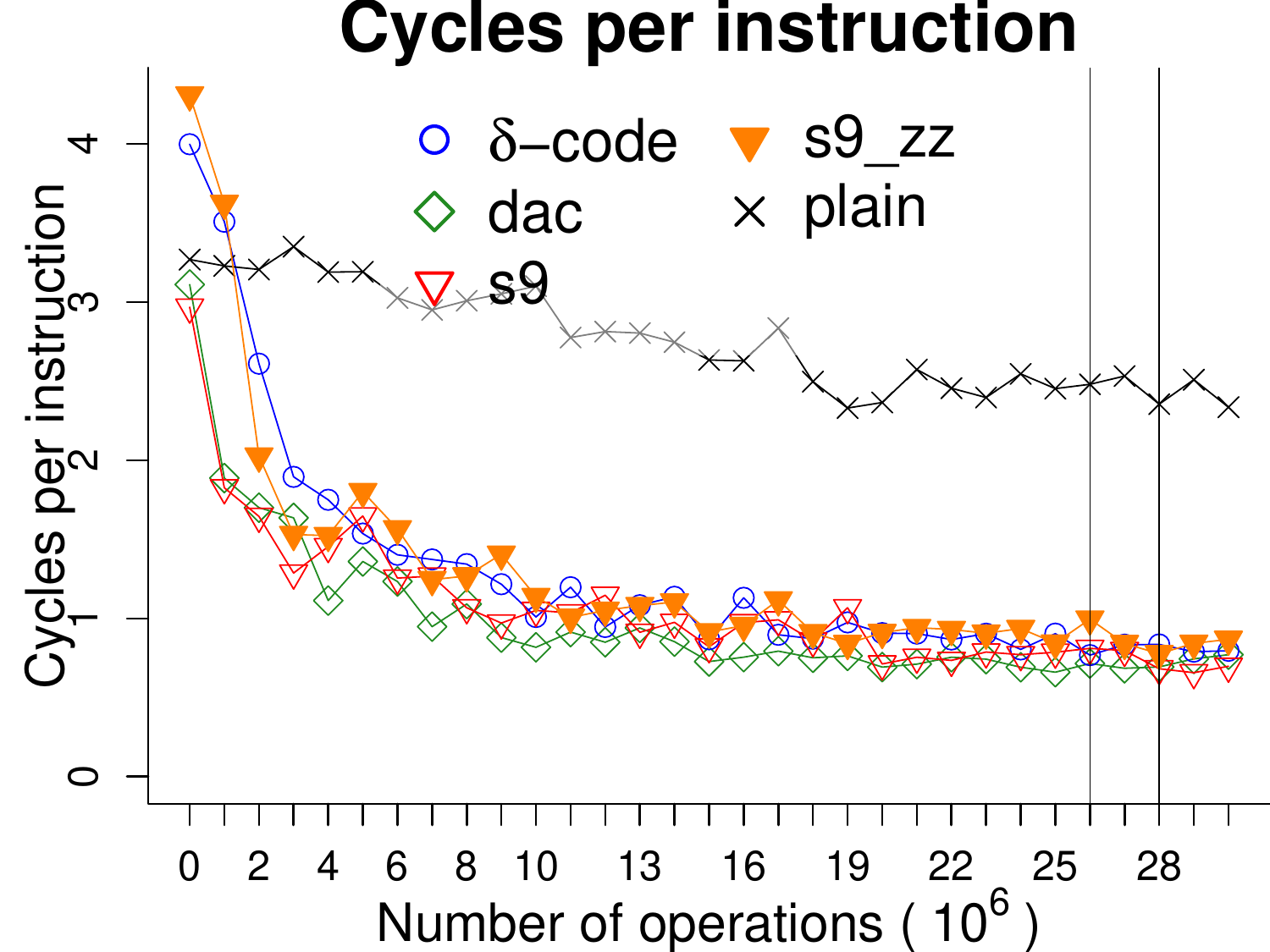}
  \end{subfigure}
  \begin{subfigure}[b]{0.32\textwidth}
    \includegraphics[width=0.985\linewidth]{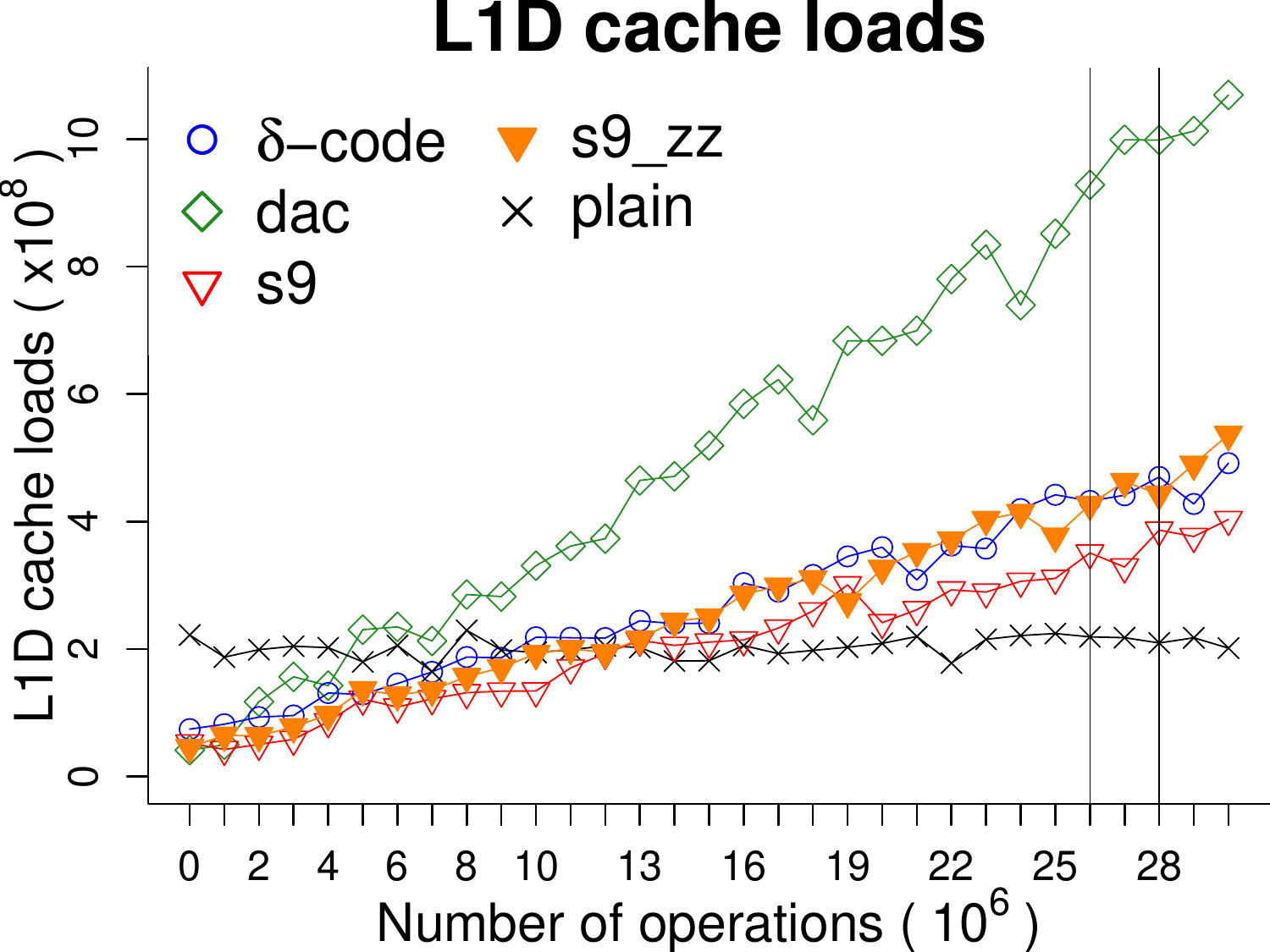}
  \end{subfigure}
  \begin{subfigure}[b]{0.32\textwidth}
    \includegraphics[width=0.985\linewidth]{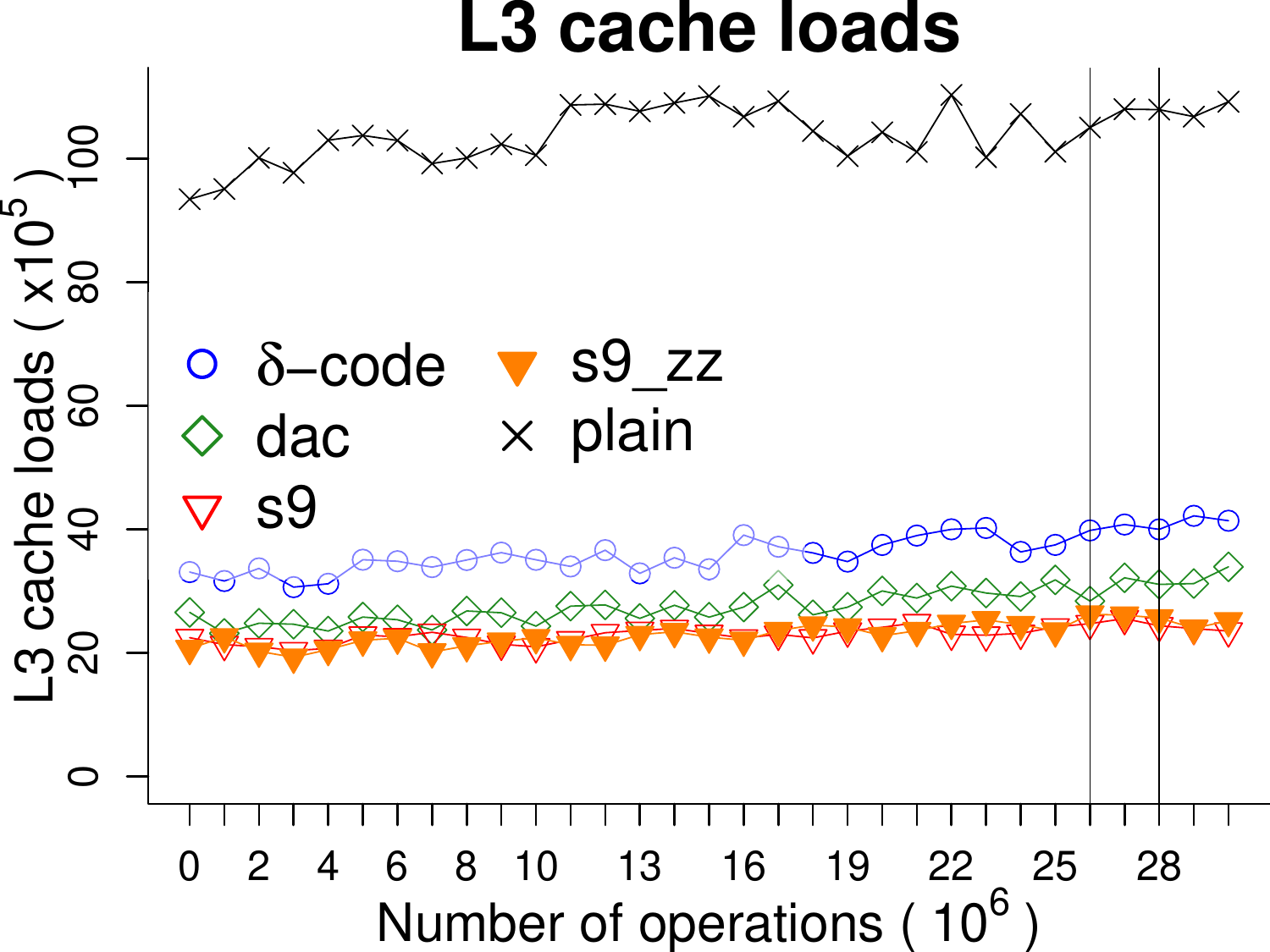}
  \end{subfigure}
  \begin{subfigure}[b]{\textwidth}
    \caption{{\tt lcp} vector of the dataset {\tt dblp}}
    \label{fig:seq-dblp-lcp}
  \end{subfigure}
  
  \caption{Experiments with sequential access pattern (Part II)}
  \label{fig:sequential_access_2}
\end{figure*}

Figures~\ref{fig:binsearch} shows the running time and the energy consumption from zero up to 24,000 binary searches over the integer vector {\tt sorted}, where each binary search performs up to $\log(104,857,600)\approx 27$ random accesses to the vector. We only report the \civ{}s whose sizes are smaller than those of their corresponding vectors in plain form. Both the running time and the energy consumption increase with the number of binary searches. When the number of binary searches is zero, the vector is loaded in memory, without accessing their values. Among of the tested \civ{}s, {\tt dac\_zz} is the one using less space but not the most efficient in running time nor energy consumption. The most time- and energy-efficient approach was the \civ{} {\tt s9\_zz}, which up to 24,000 binary searches uses less time and energy than the vector in plain form. Thus, as a preliminary conclusion, \civ{}s represent an energy-efficient alternative for binary search when the number of binary searches executed over a vector is limited: less than 4,000 binary searches for {\tt $\delta$-code\_zz}, {\tt $\gamma$-code\_zz} and {\tt dac\_zz}, and less than 24,000 for {\tt s9\_zz}. 
We provide a deeper discussion of the factors that affect the energy consumption of \civ{}s in the following paragraphs, when we study the sequential and random access patterns.

\begin{figure*}[t]
  \centering
  \begin{subfigure}[b]{0.33\textwidth}
    \includegraphics[width=0.985\linewidth]{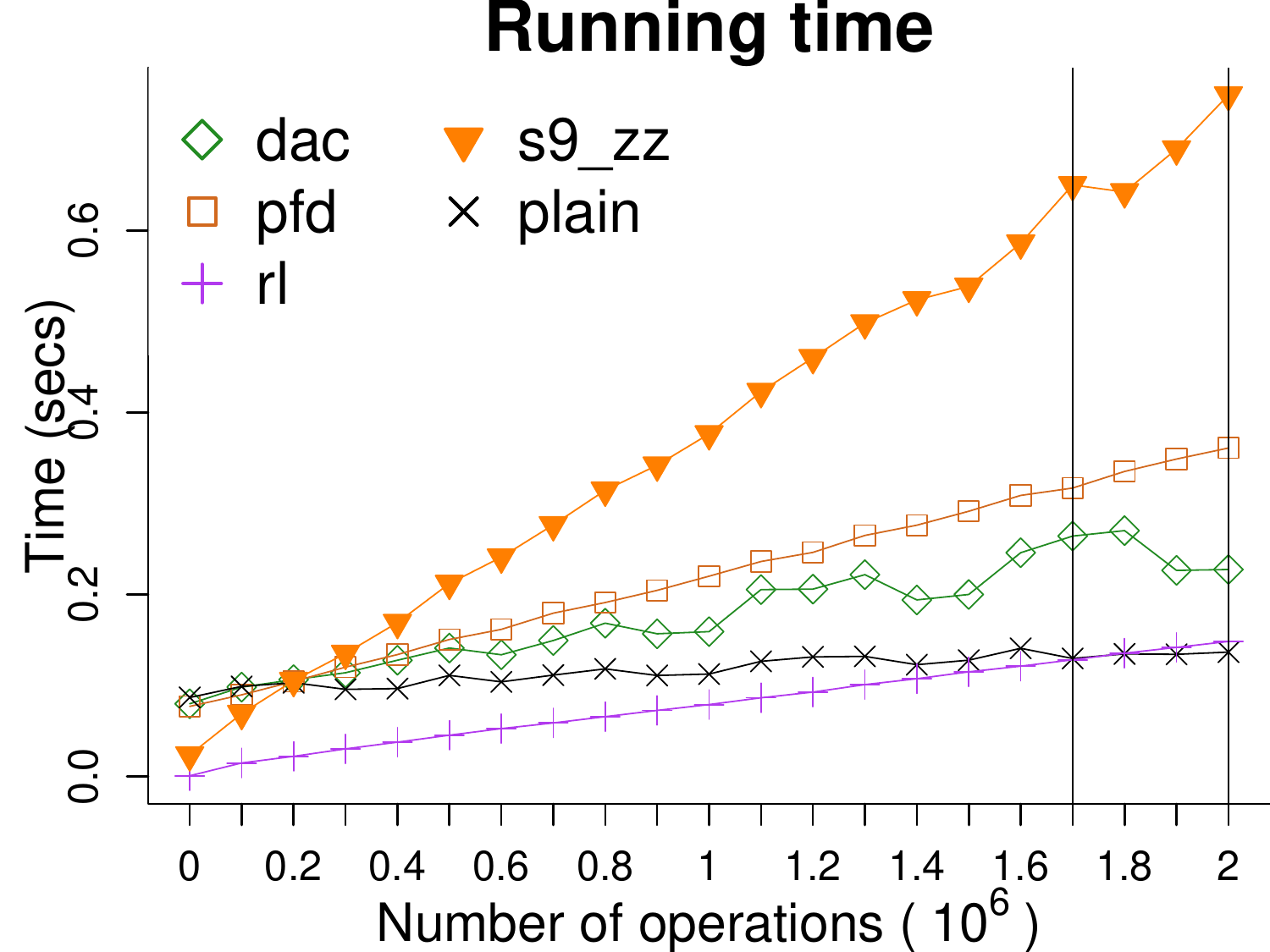}
  \end{subfigure}\hspace{0em}%
  \begin{subfigure}[b]{0.33\textwidth}
    \includegraphics[width=0.985\linewidth]{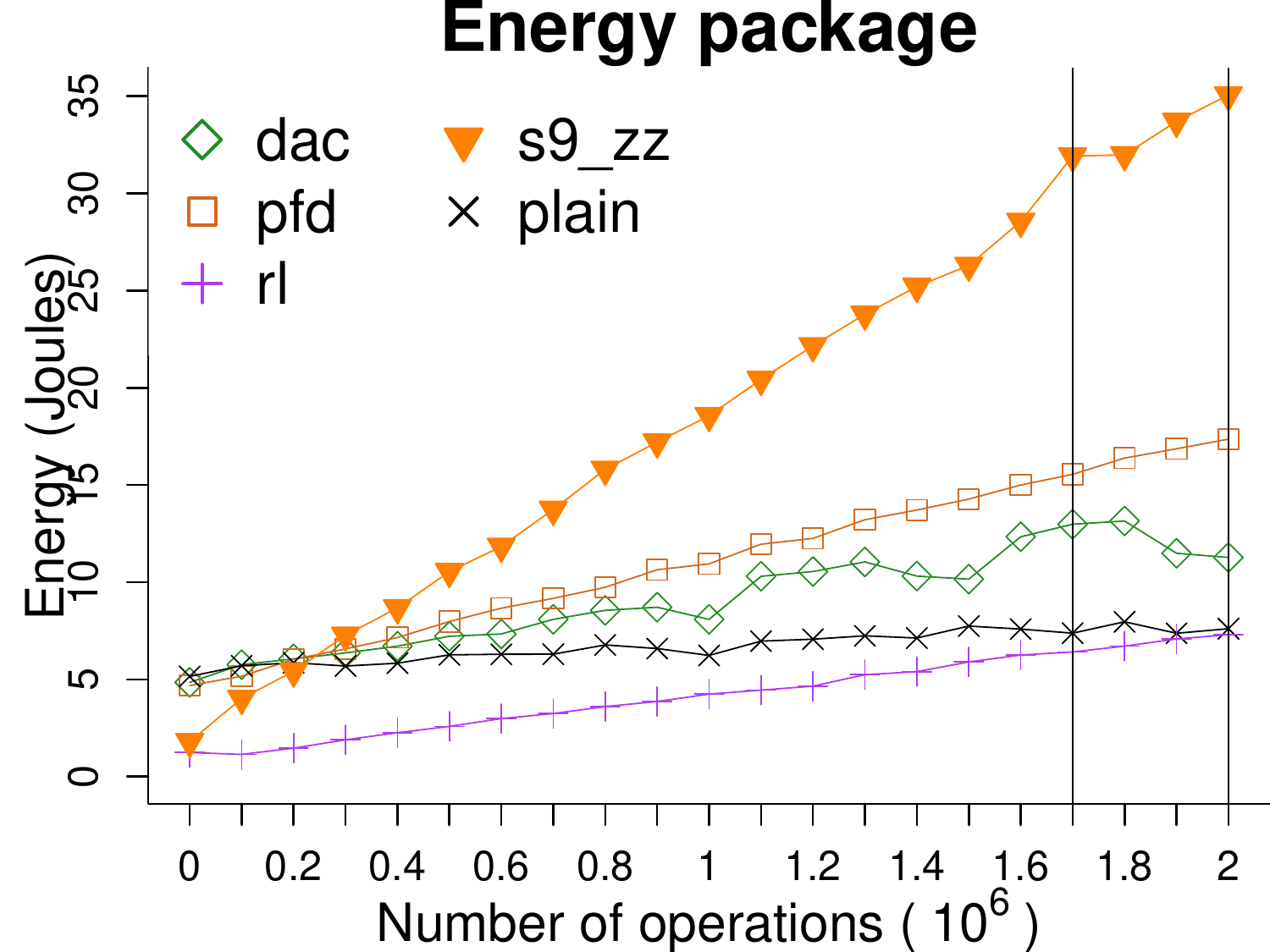}
  \end{subfigure}\hspace{0em}%
  \begin{subfigure}[b]{0.33\textwidth}
    \includegraphics[width=0.985\linewidth]{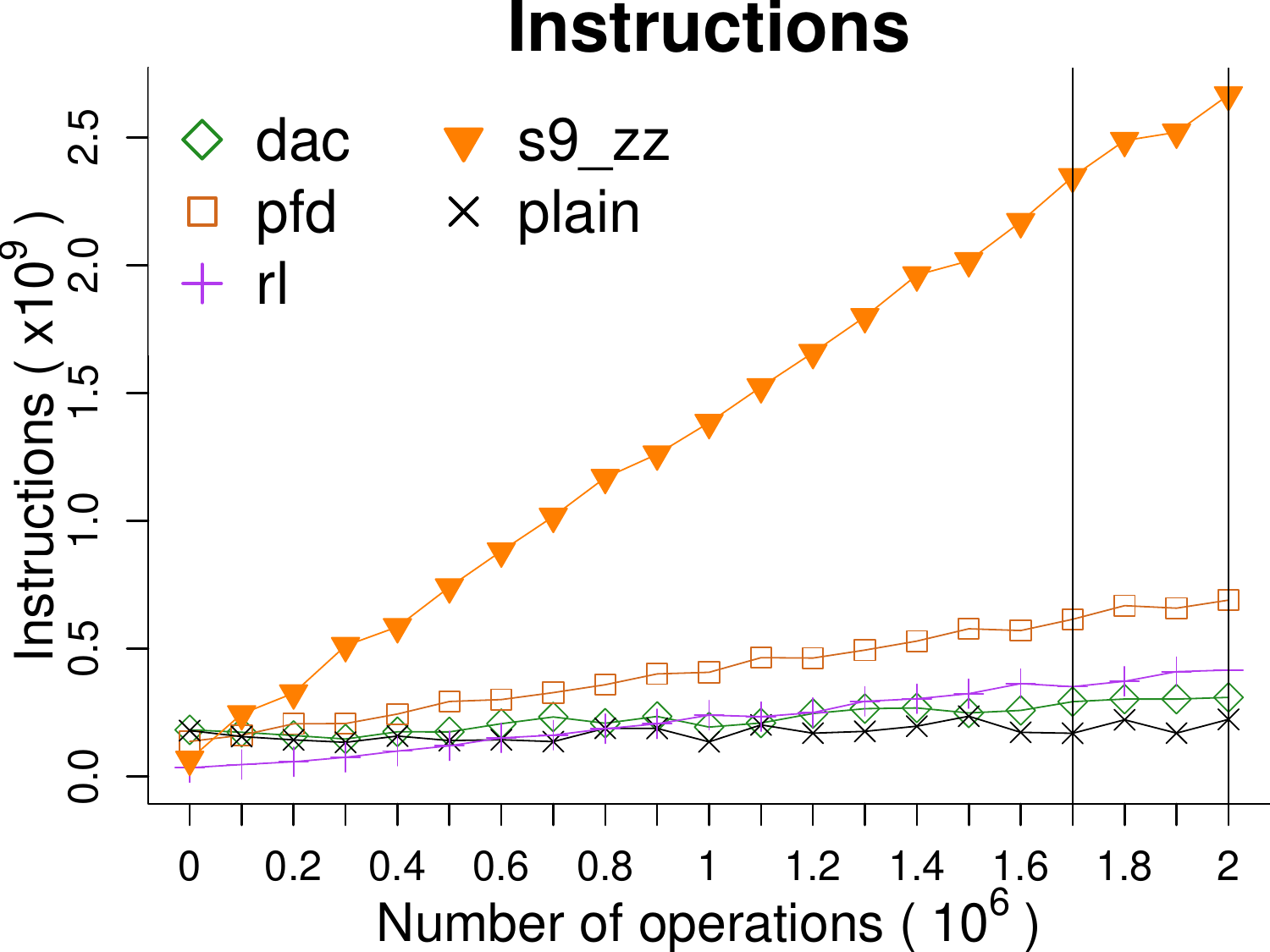}
  \end{subfigure}
  \begin{subfigure}[b]{0.32\textwidth}
    \includegraphics[width=0.985\linewidth]{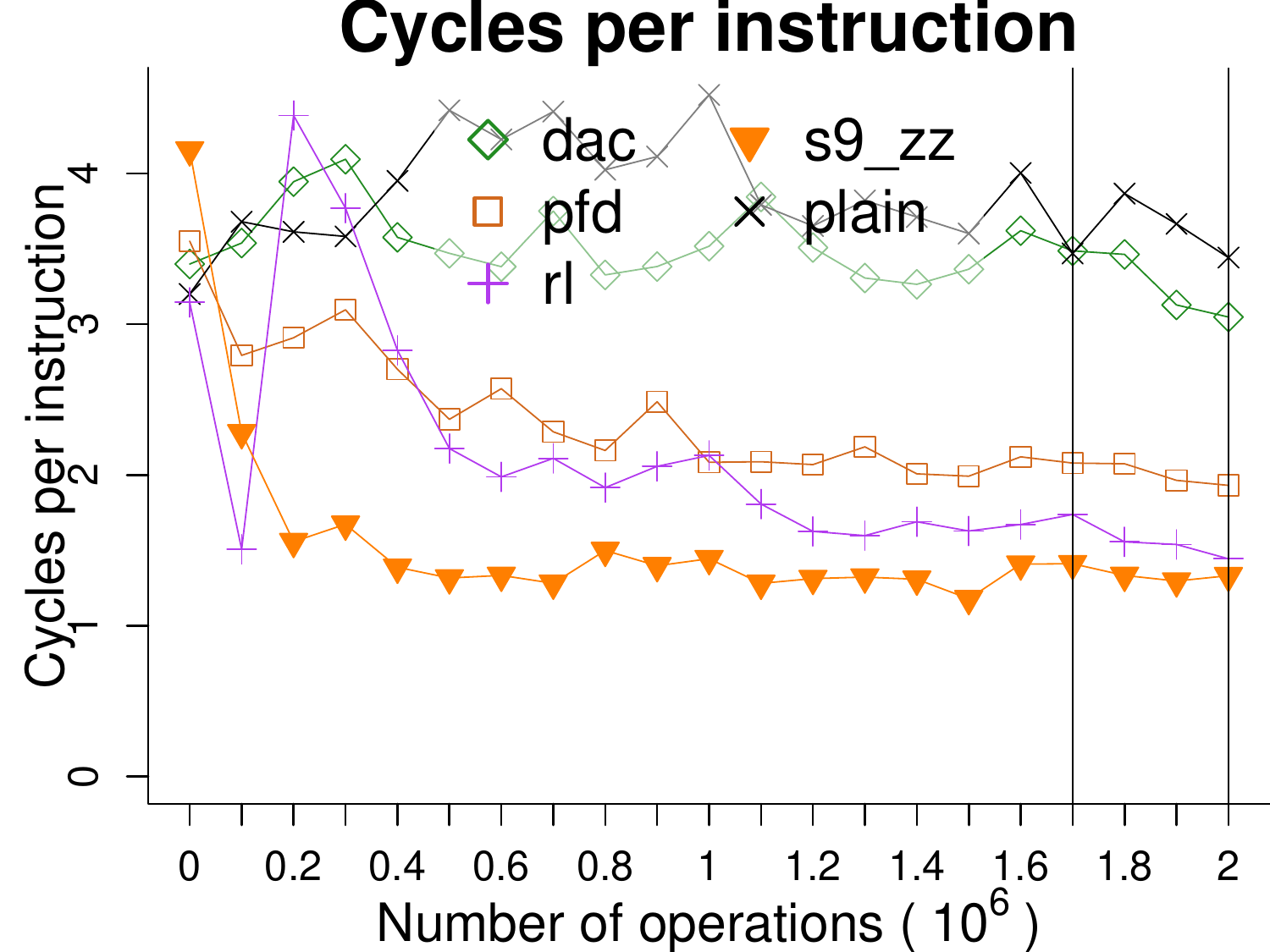}
  \end{subfigure}
  \begin{subfigure}[b]{0.32\textwidth}
    \includegraphics[width=0.985\linewidth]{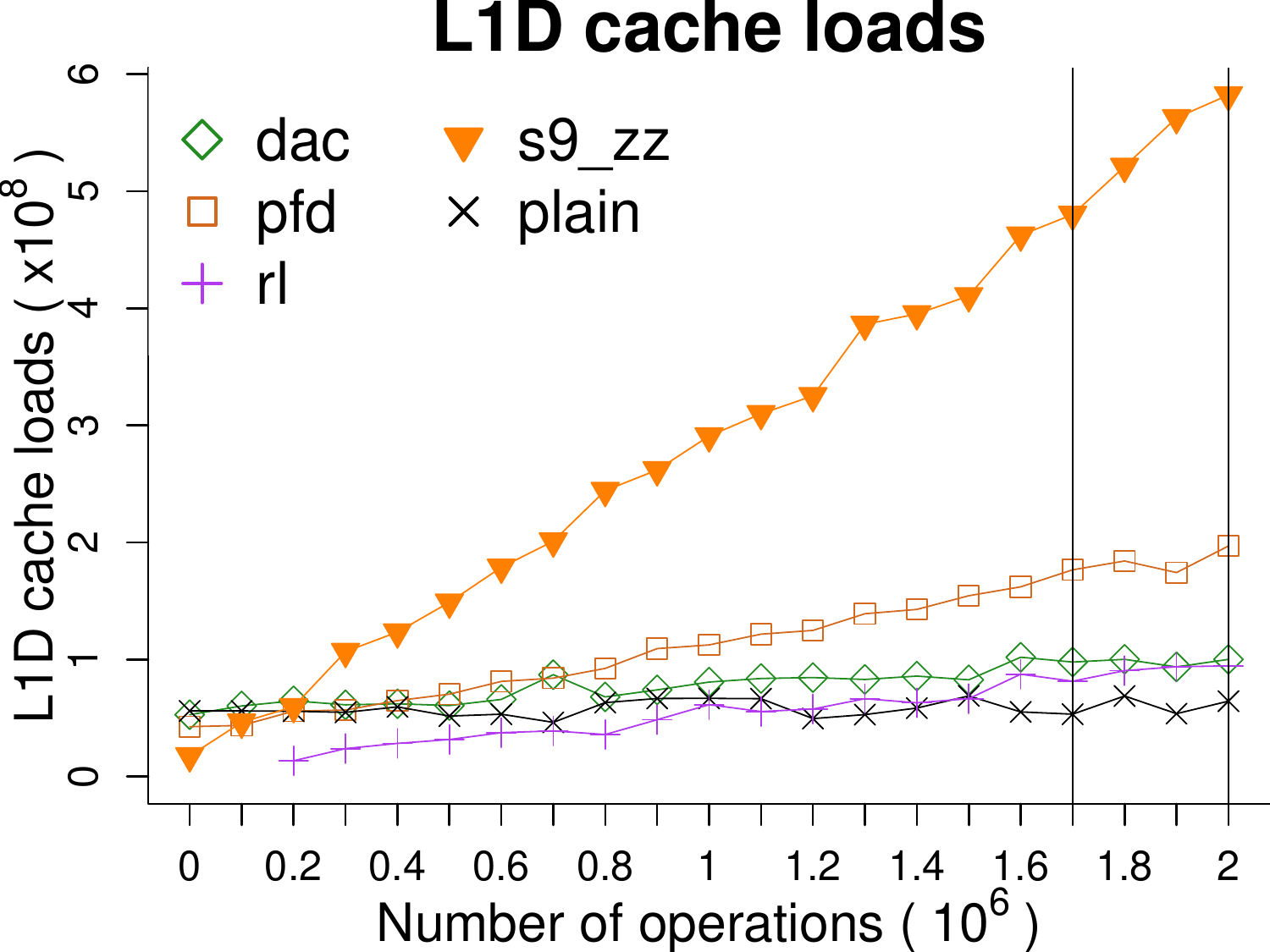}
  \end{subfigure}
  \begin{subfigure}[b]{0.32\textwidth}
    \includegraphics[width=0.985\linewidth]{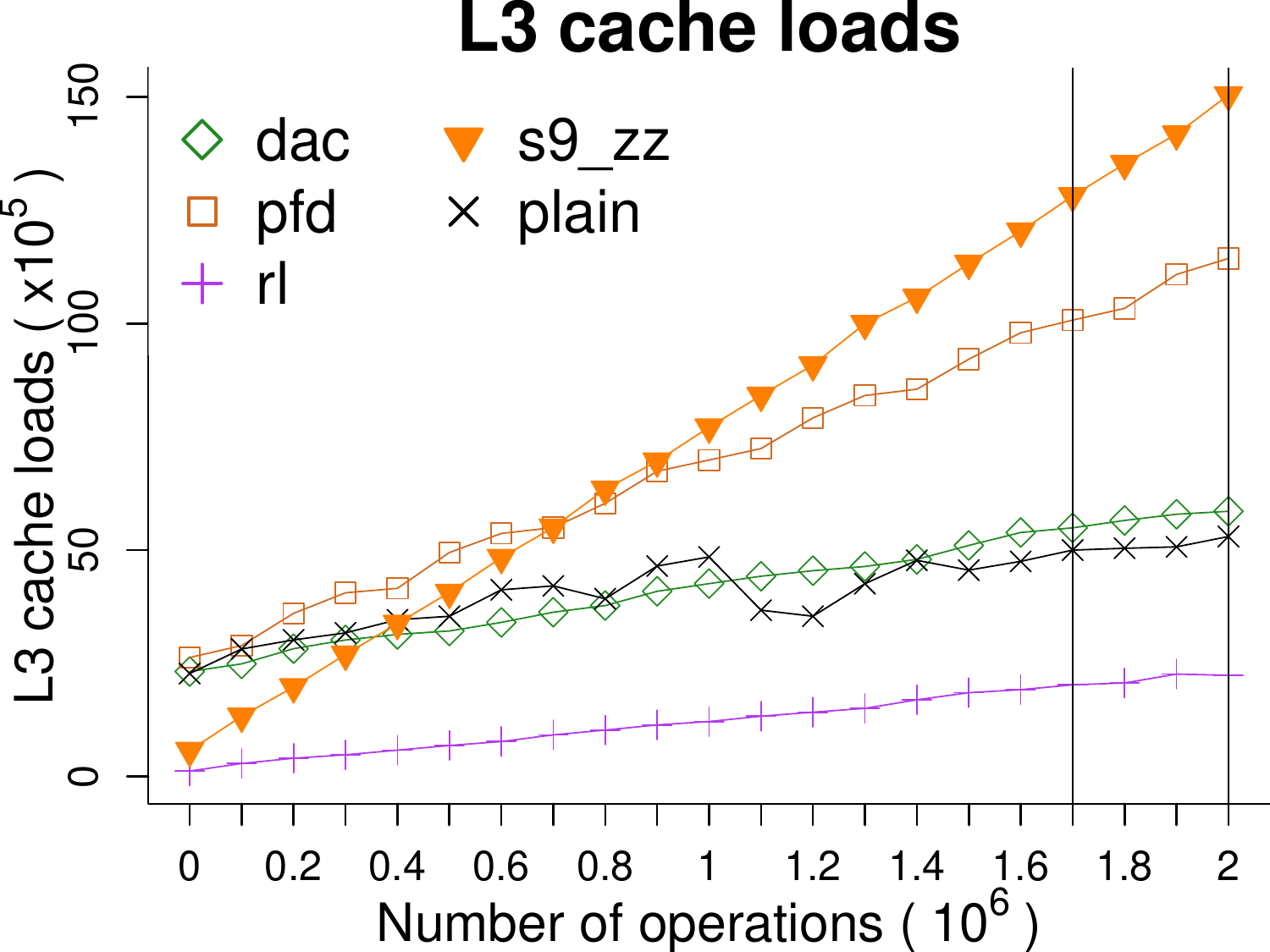}
  \end{subfigure}
  \begin{subfigure}[b]{\textwidth}
    \caption{{\tt bwt} vector of the dataset {\tt eins}}
    \label{fig:rand-bwt-eins}
  \end{subfigure}
  \begin{subfigure}[b]{0.33\textwidth}
    \includegraphics[width=0.985\linewidth]{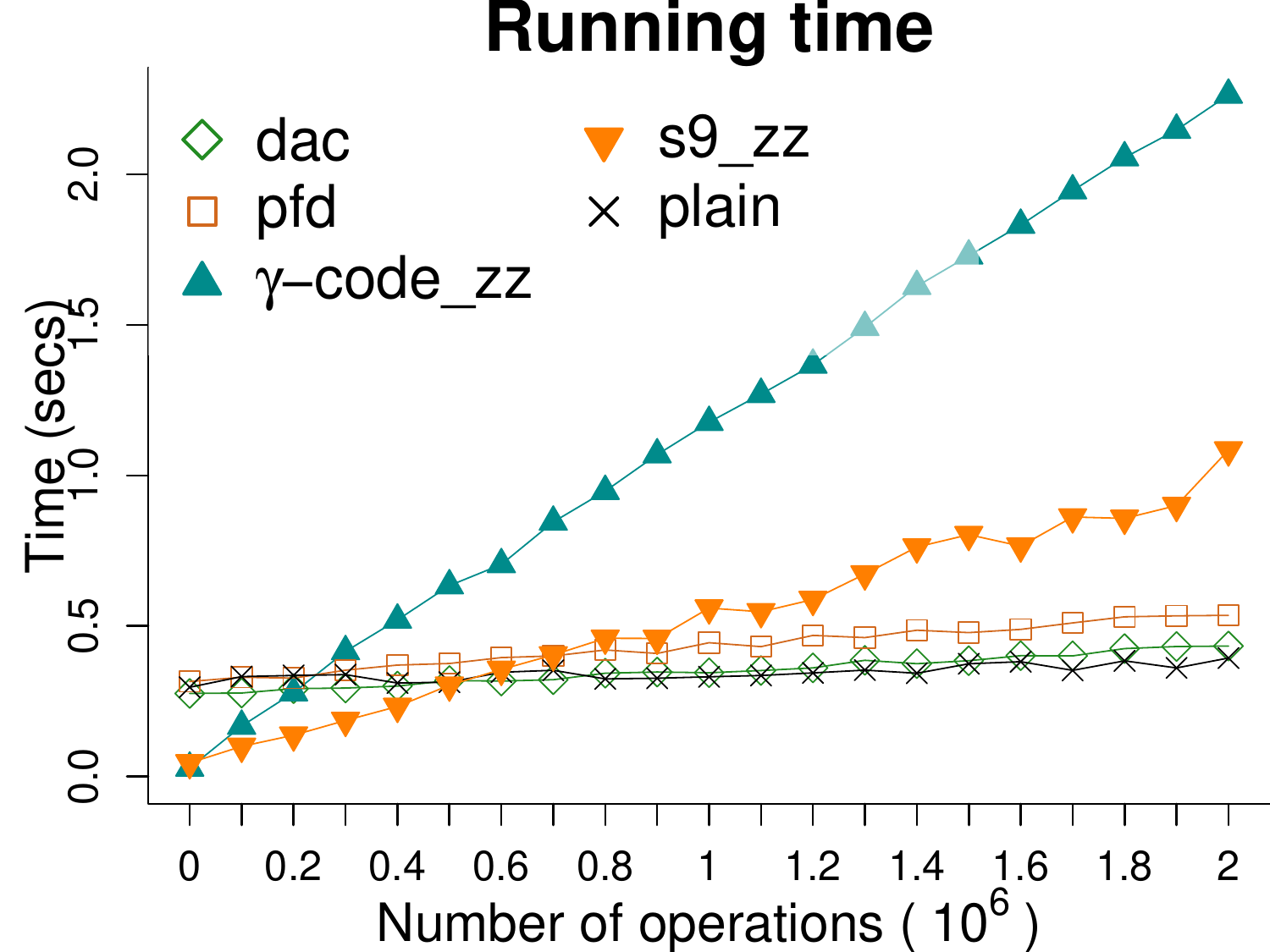}
  \end{subfigure}\hspace{0em}%
  \begin{subfigure}[b]{0.33\textwidth}
    \includegraphics[width=0.985\linewidth]{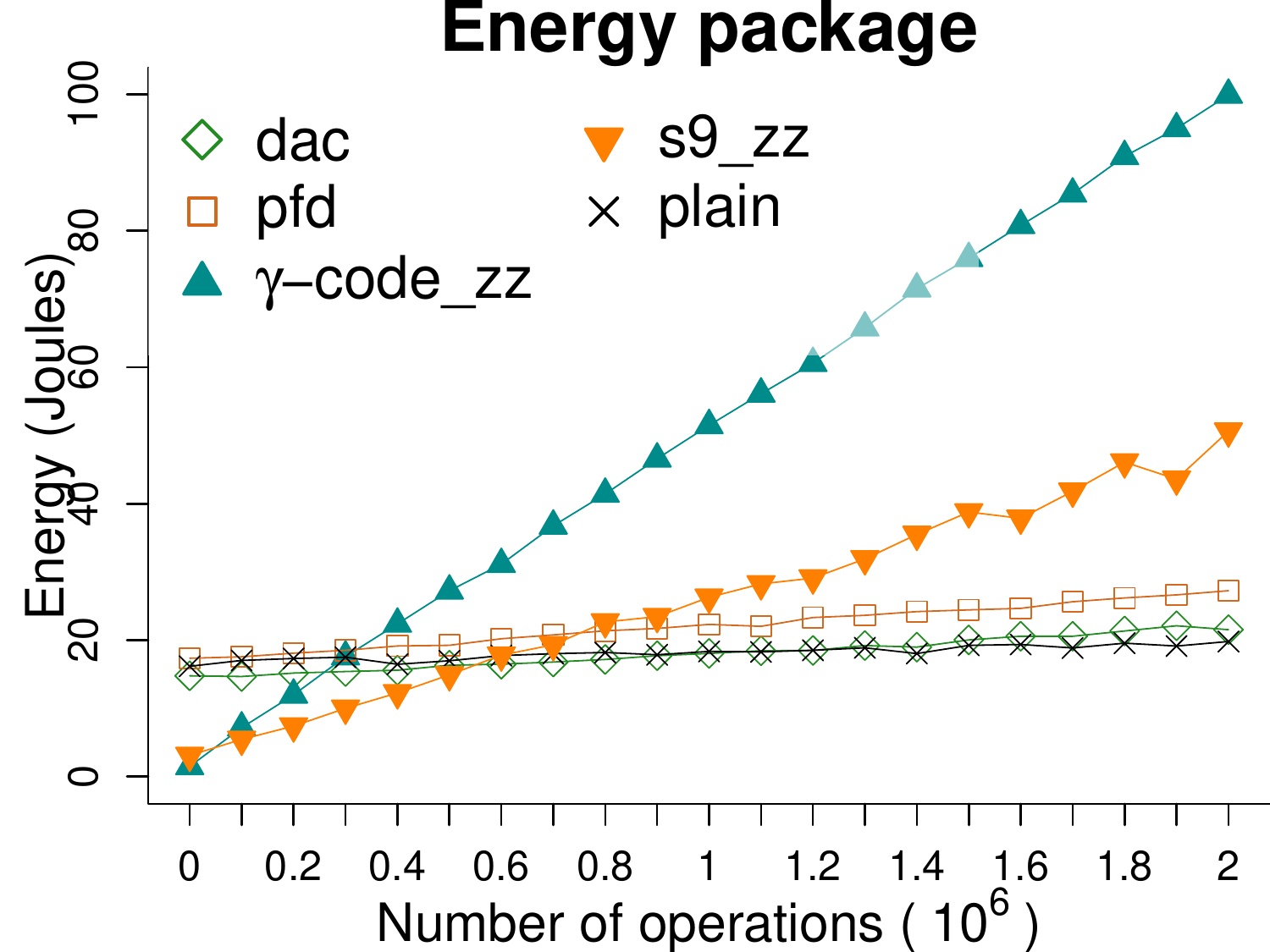}
  \end{subfigure}\hspace{0em}%
  \begin{subfigure}[b]{0.33\textwidth}
    \includegraphics[width=0.985\linewidth]{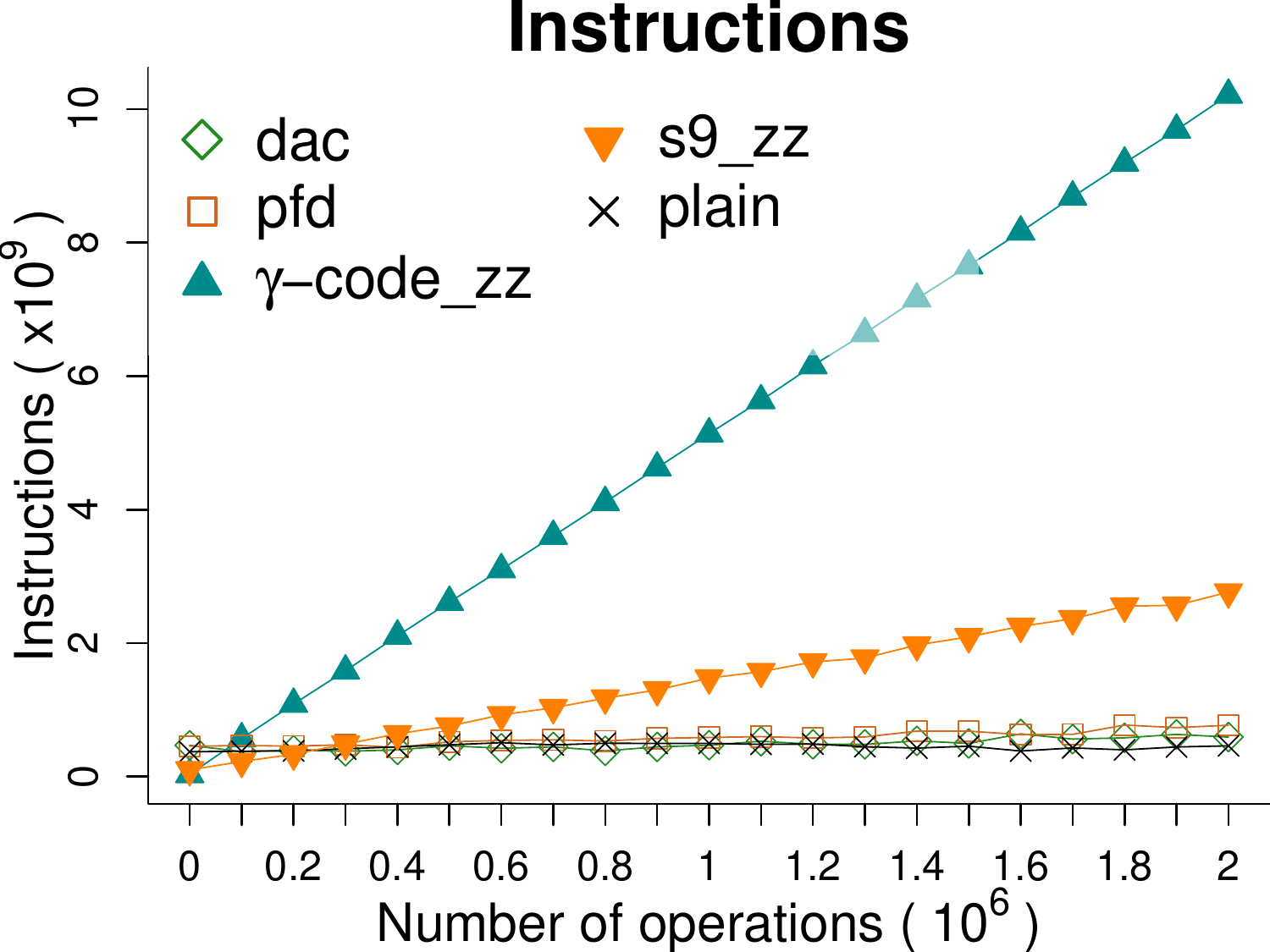}
  \end{subfigure}
  \begin{subfigure}[b]{0.32\textwidth}
    \includegraphics[width=0.985\linewidth]{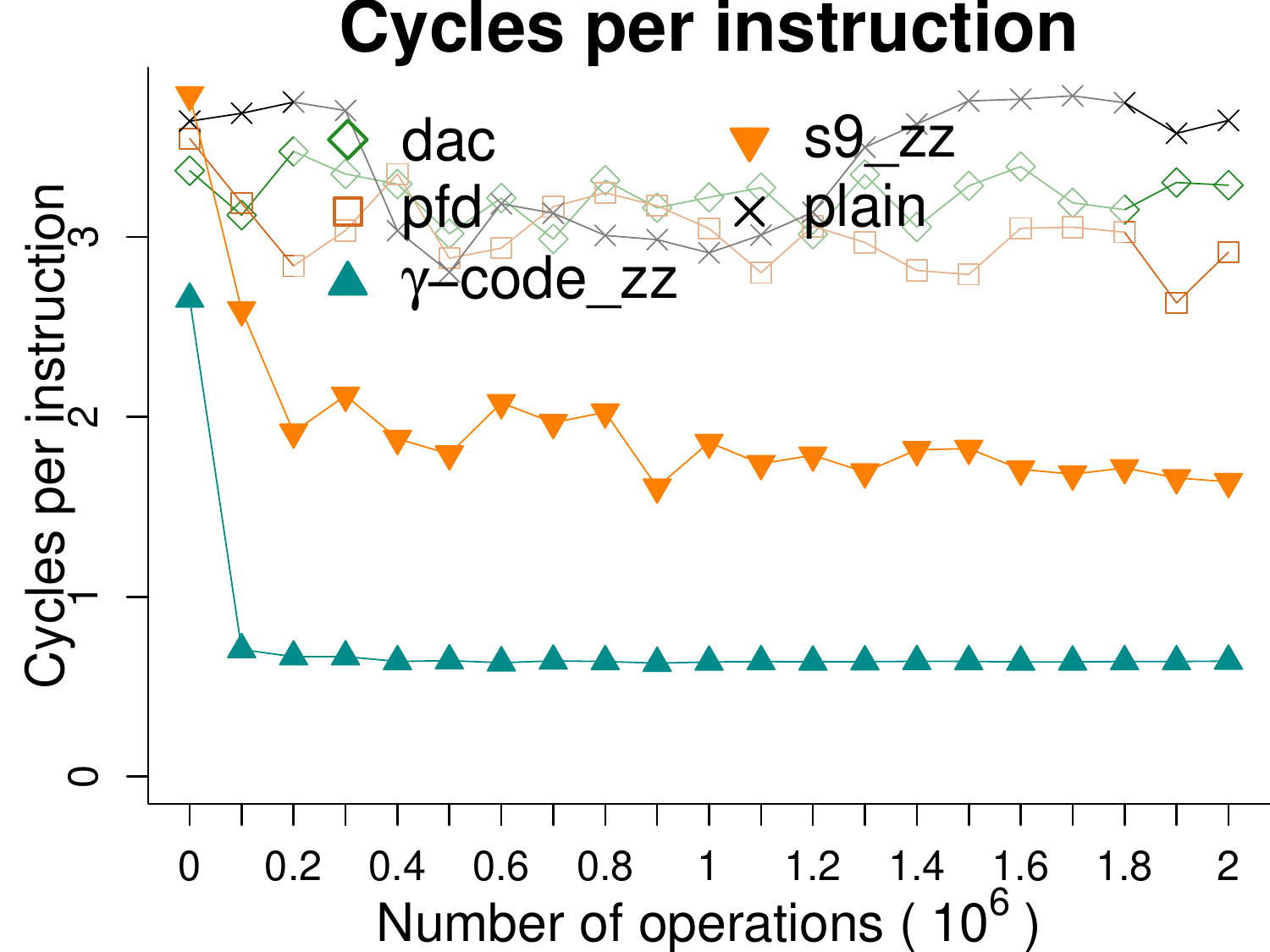}
  \end{subfigure}
  \begin{subfigure}[b]{0.32\textwidth}
    \includegraphics[width=0.985\linewidth]{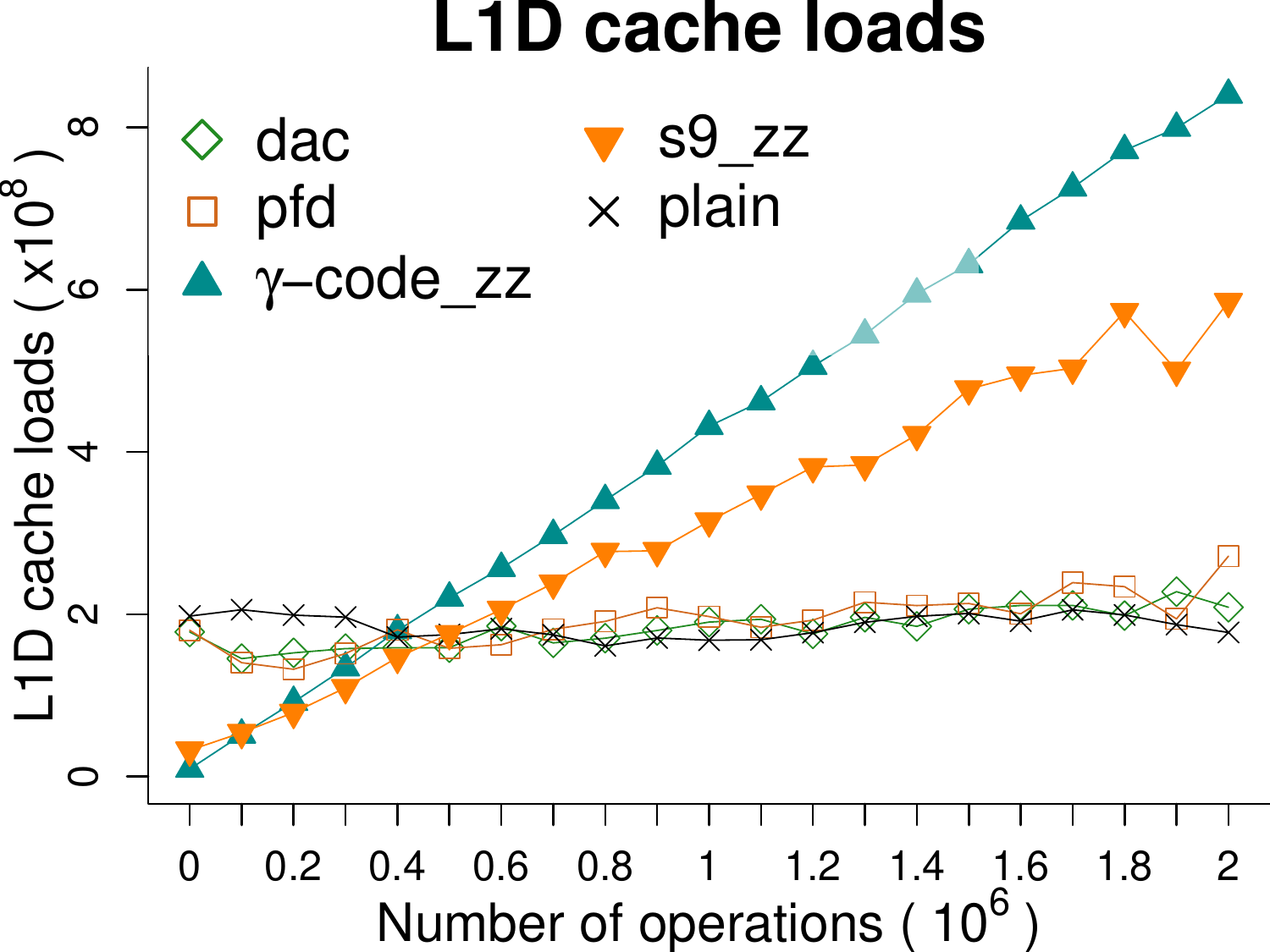}
  \end{subfigure}
  \begin{subfigure}[b]{0.32\textwidth}
    \includegraphics[width=0.985\linewidth]{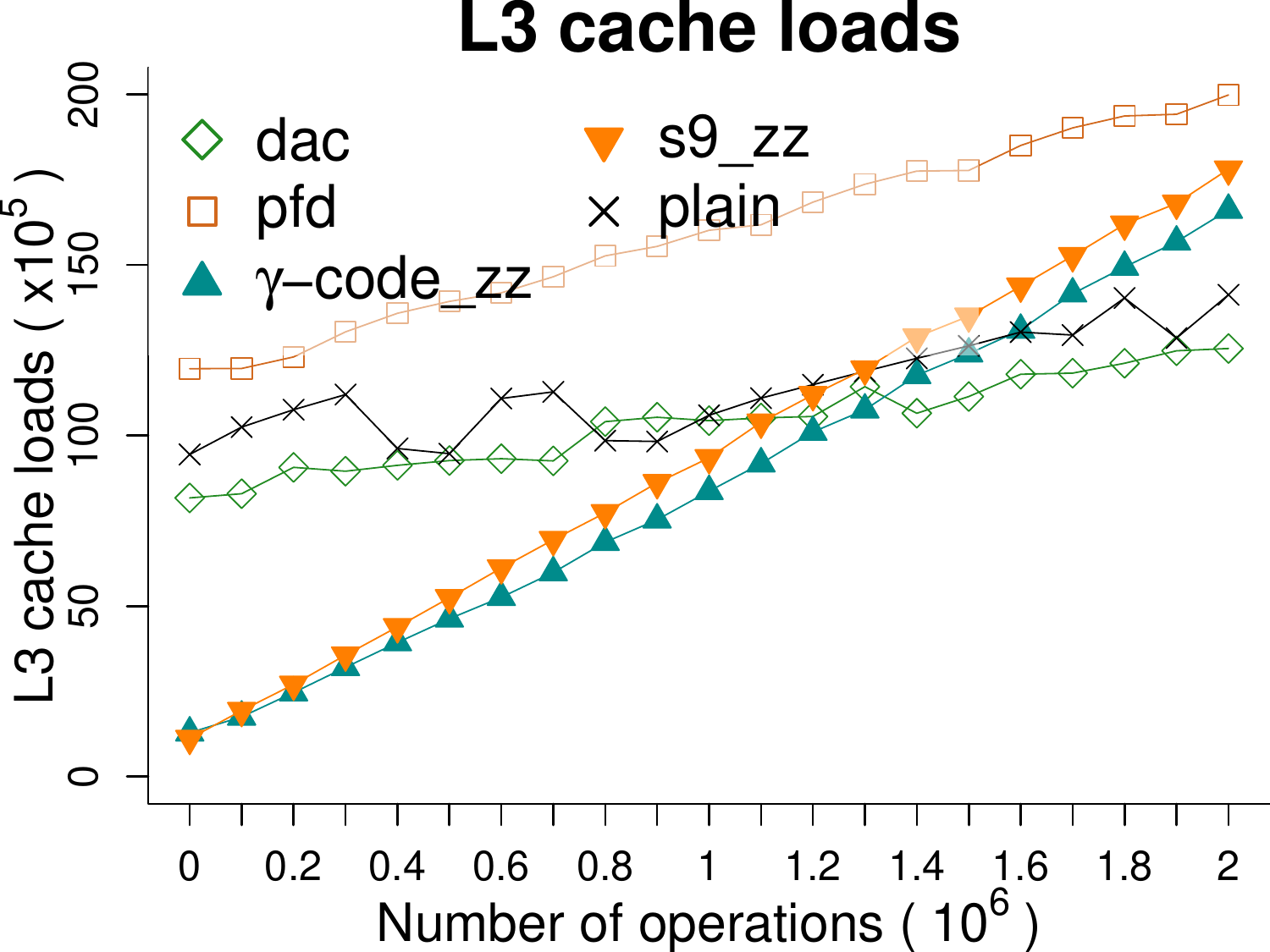}
  \end{subfigure}
  \begin{subfigure}[b]{\textwidth}
    \caption{{\tt psi} vector of the dataset {\tt kernel}}
    \label{fig:rand-kernel-psi}
  \end{subfigure}
  
  \caption{Experiments with random access pattern (Part I)}
  \label{fig:random_access_1}
\end{figure*}

\begin{figure*}[t]
  \centering
  \begin{subfigure}[b]{0.33\textwidth}
    \includegraphics[width=0.985\linewidth]{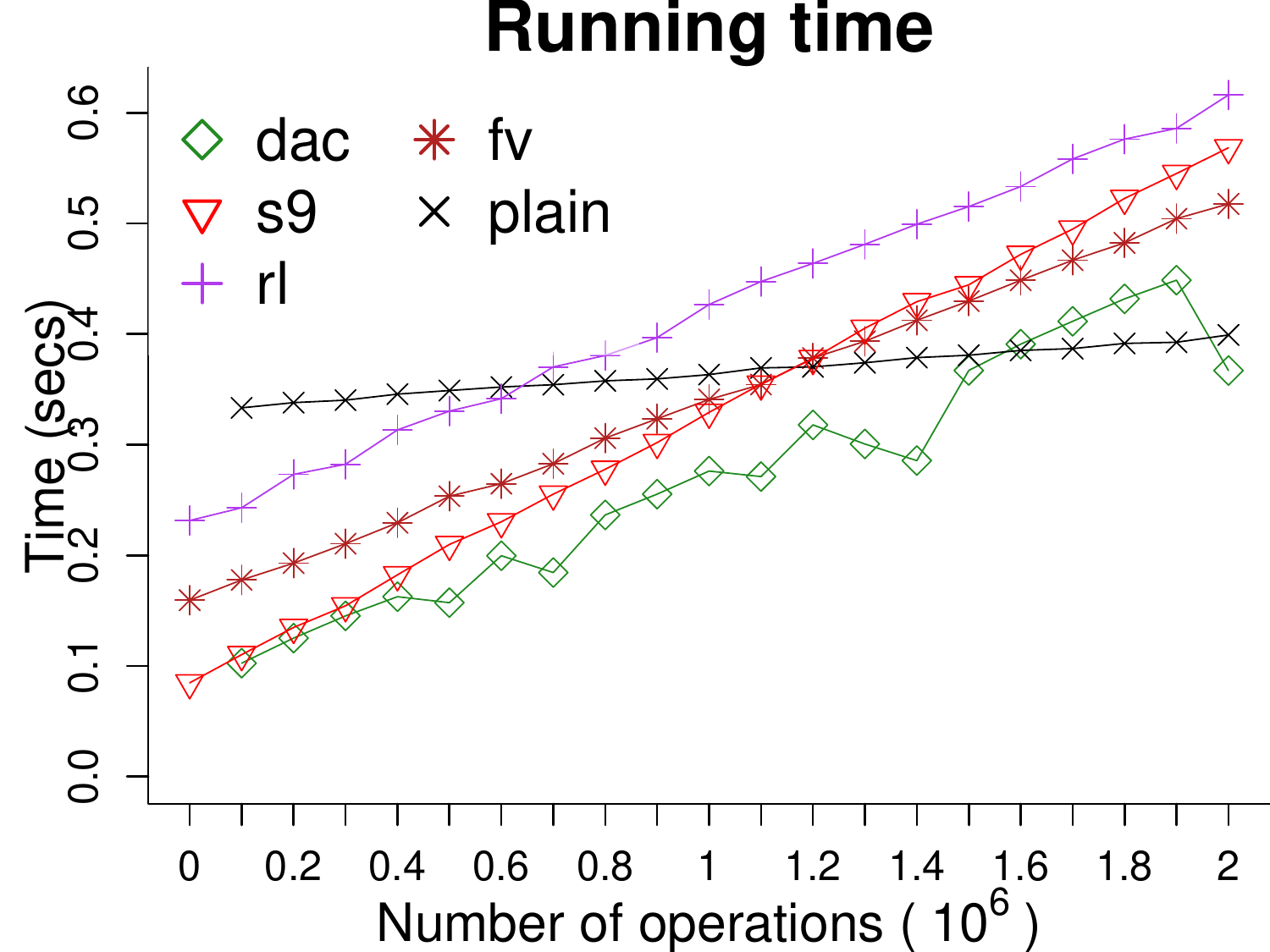}
  \end{subfigure}\hspace{0em}%
  \begin{subfigure}[b]{0.33\textwidth}
    \includegraphics[width=0.985\linewidth]{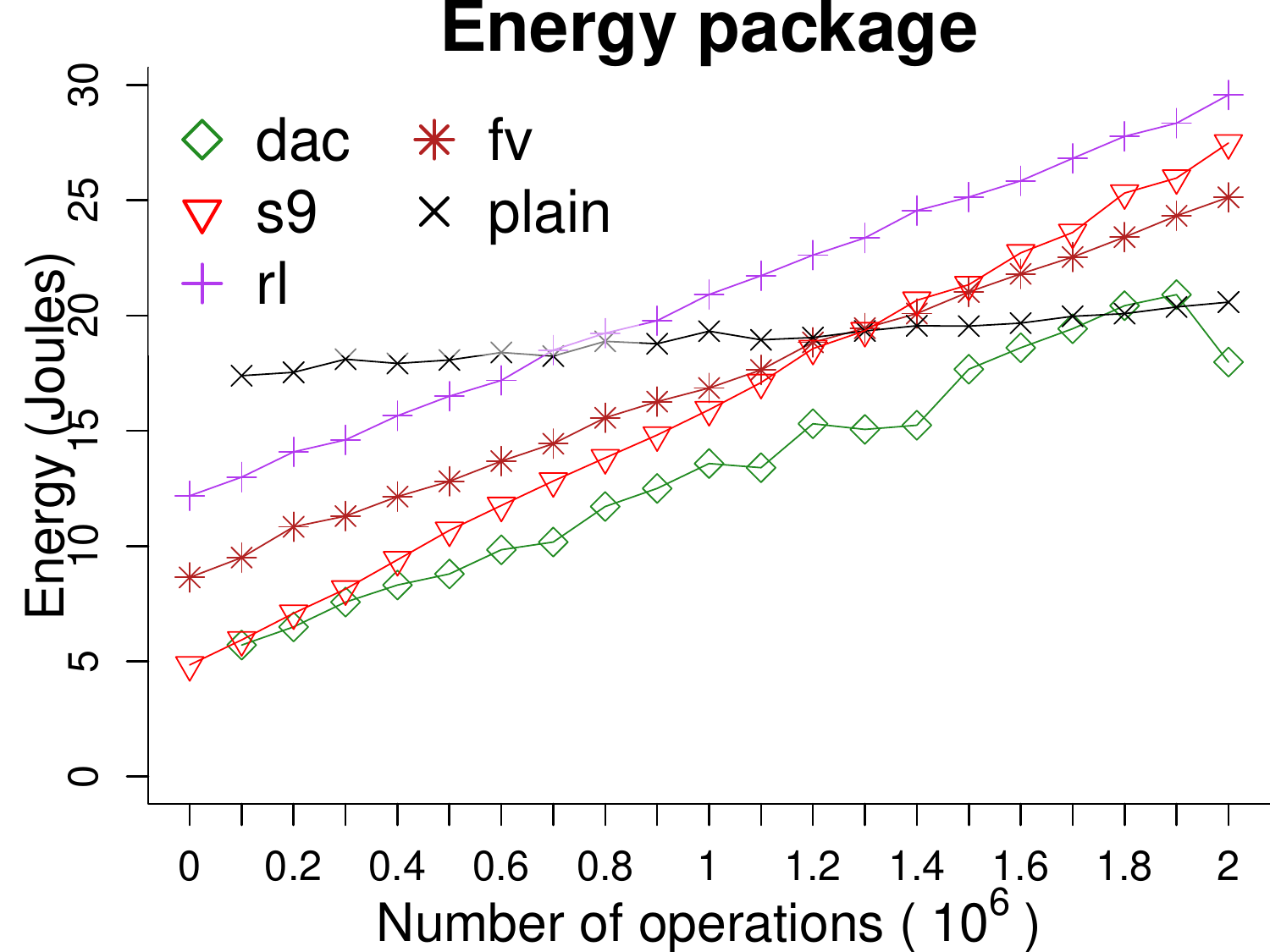}
  \end{subfigure}\hspace{0em}%
  \begin{subfigure}[b]{0.33\textwidth}
    \includegraphics[width=0.985\linewidth]{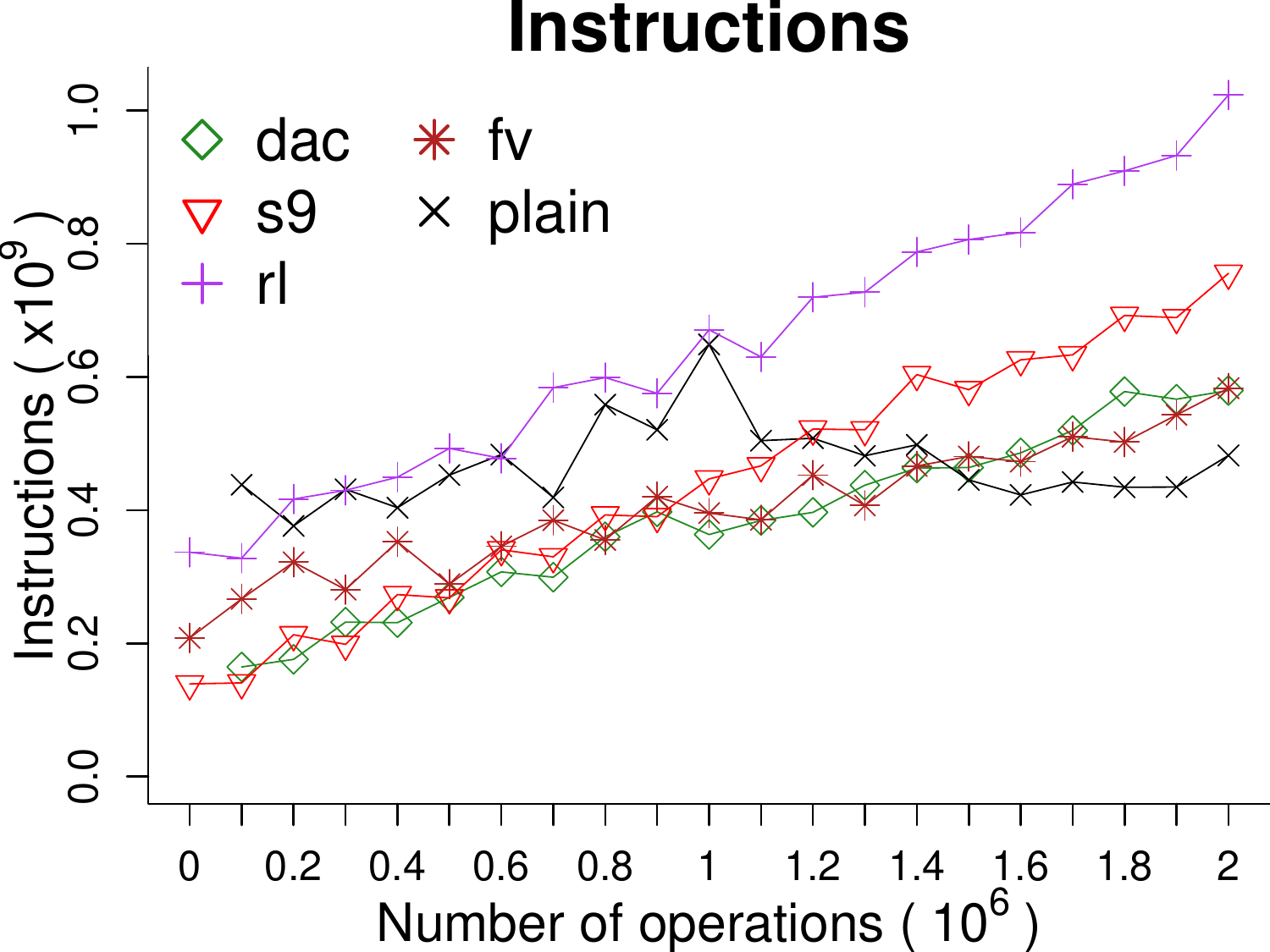}
  \end{subfigure}
  \begin{subfigure}[b]{0.32\textwidth}
    \includegraphics[width=0.985\linewidth]{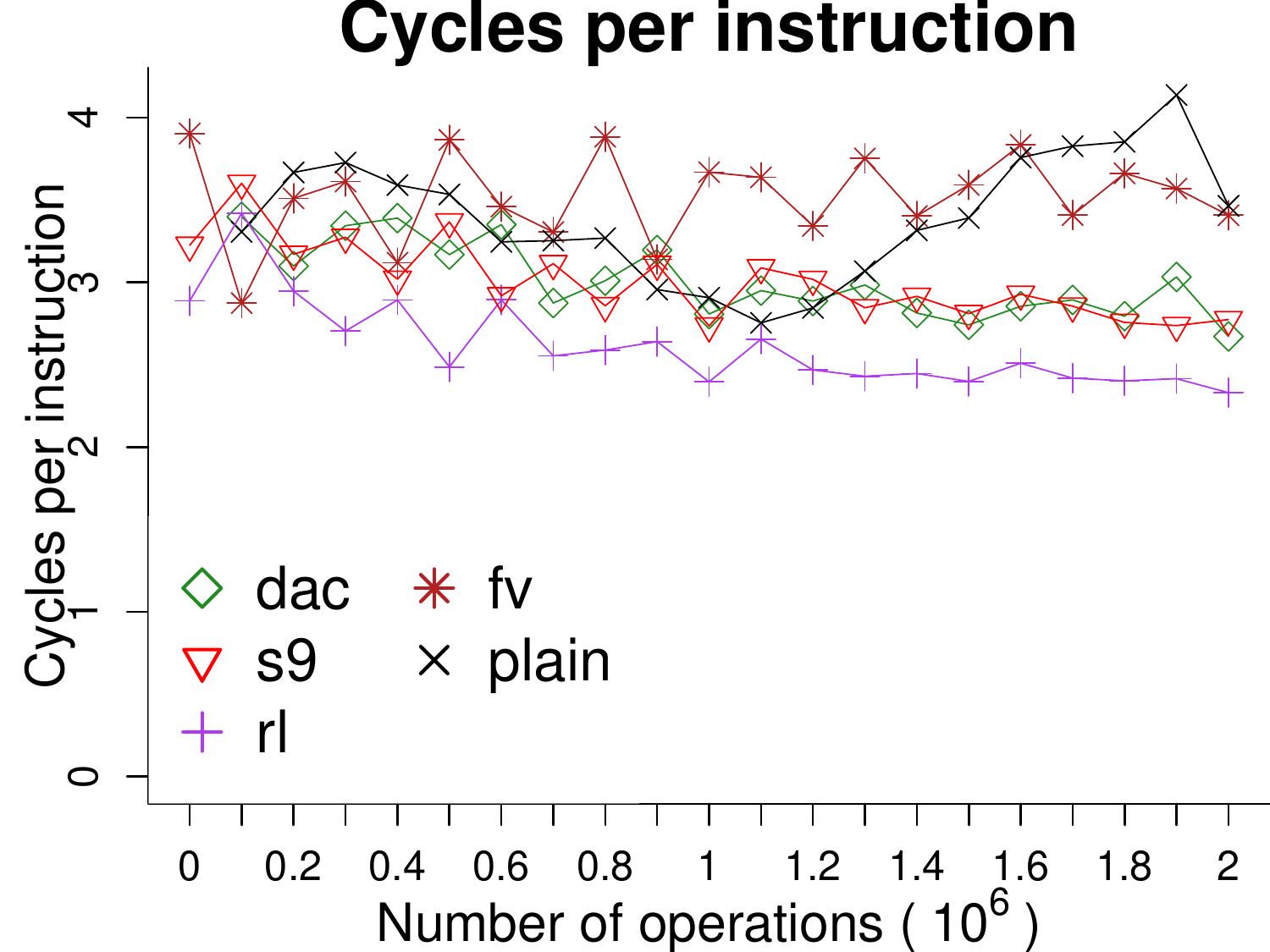}
  \end{subfigure}
  \begin{subfigure}[b]{0.32\textwidth}
    \includegraphics[width=0.985\linewidth]{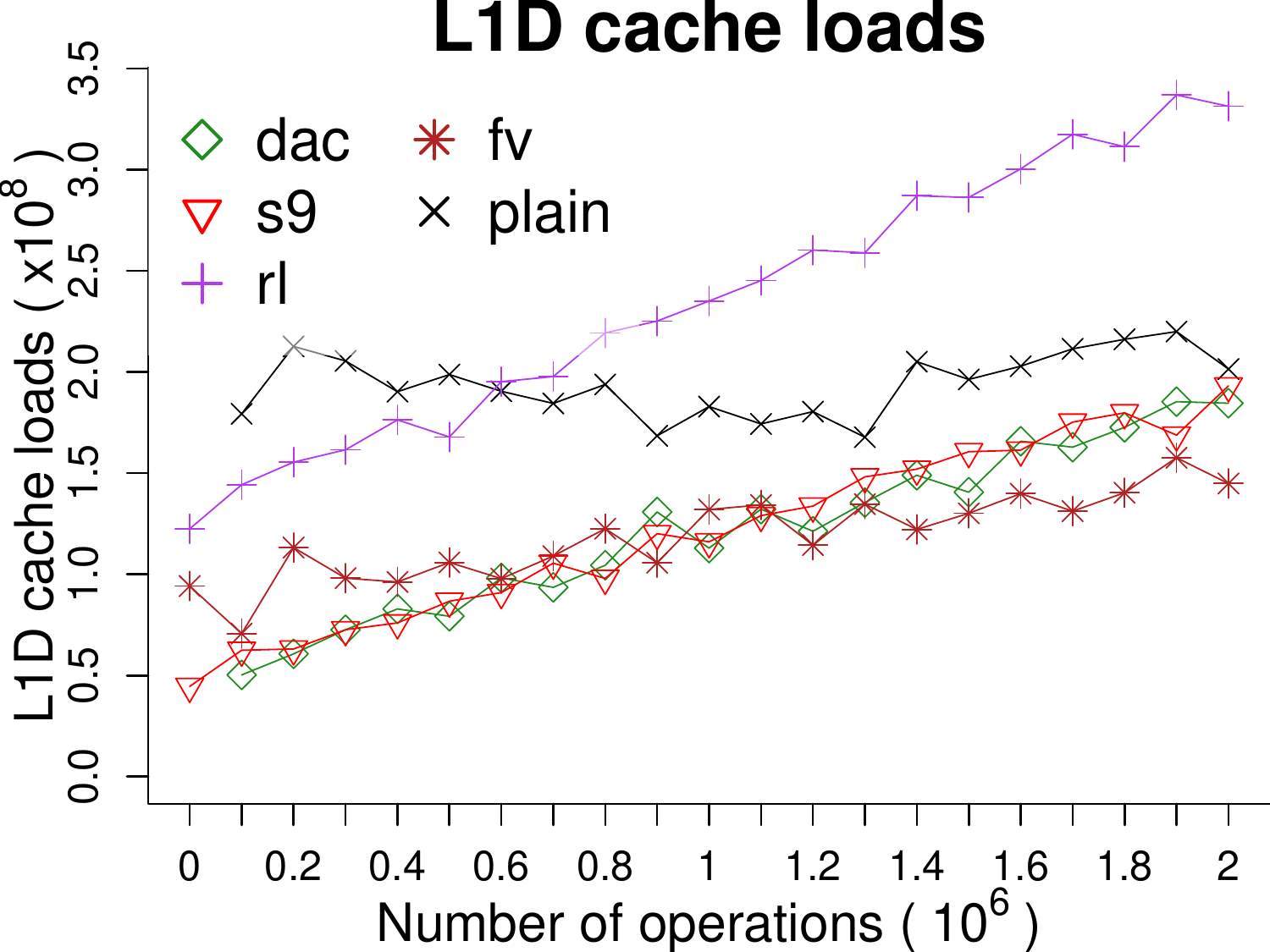}
  \end{subfigure}
  \begin{subfigure}[b]{0.32\textwidth}
    \includegraphics[width=0.985\linewidth]{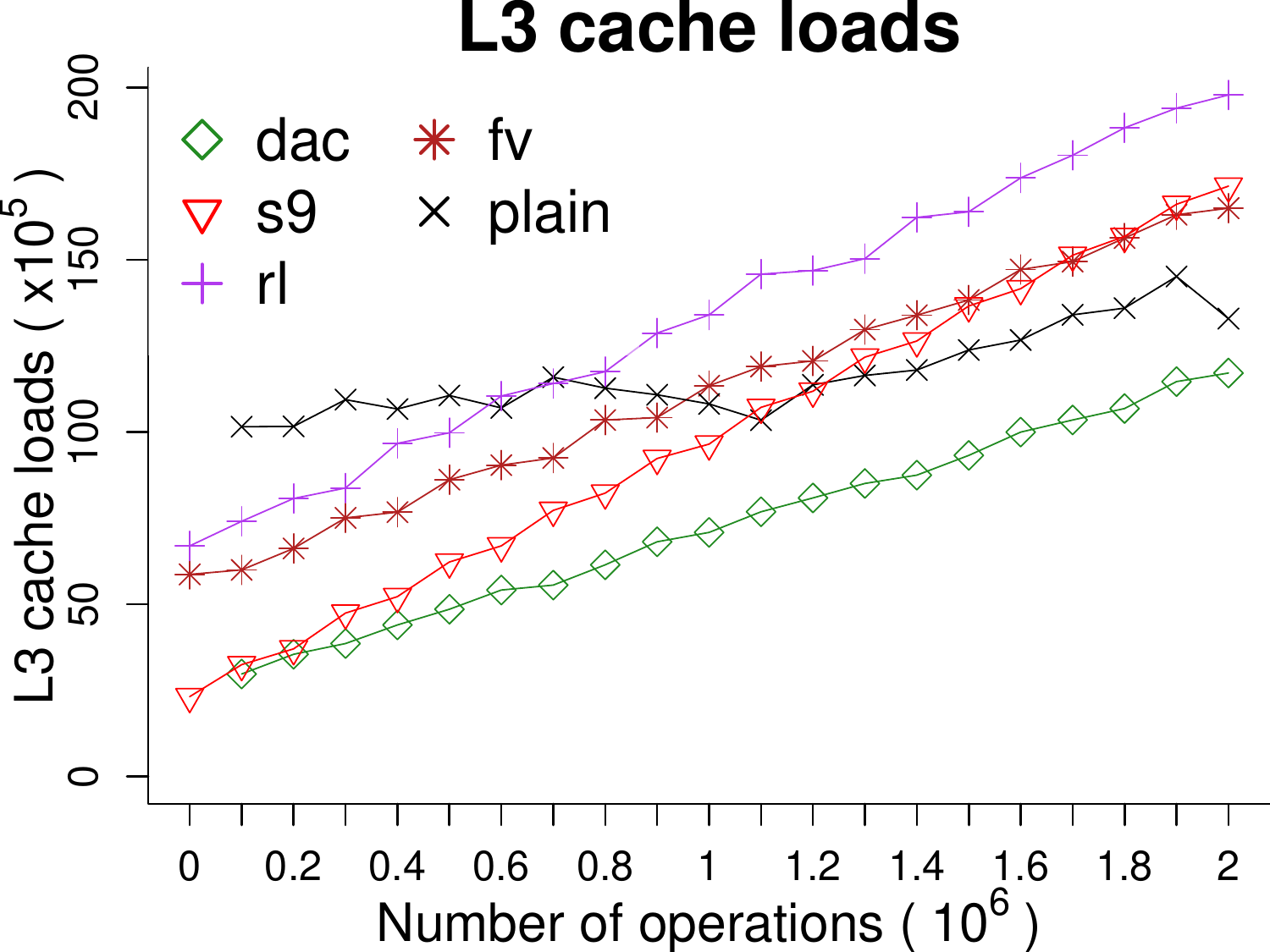}
  \end{subfigure}
  \begin{subfigure}[b]{\textwidth}
    \caption{{\tt lcp} vector of the dataset {\tt dblp}}
    \label{fig:rand-dblp-lcp}
  \end{subfigure}
  
  \caption{Experiments with random access pattern (Part II)}
  \label{fig:random_access_2}
\end{figure*}

Figures~\ref{fig:sequential_access_1}-\ref{fig:random_access_2} show several
metrics for some of the datasets. We selected three
representative cases to discuss our findings: For {\tt bwt}, {\tt psi} and {\tt
  lcp} vectors, the results of the {\tt eins}, {\tt kernel} and {\tt dblp}
datasets are shown, respectively. We omitted experiments for some \civ{}s that
use more space than the vectors in plain format, such as the case of {\tt fv}
for the {\tt bwt} vector of the dataset {\tt cere}, among others. To improve the readability of the figures, we only show the five \civ{}s with better energy consumption for each experiment.

As in Figure~\ref{fig:binsearch}, when the number
of operations is zero, the \civ{}s are loaded in memory, without accessing their values. 
Figures~\ref{fig:sequential_access_1} and \ref{fig:sequential_access_2} show the results for sequential access
pattern. For the {\tt bwt} vector of {\tt eins}, in
Figure~\ref{fig:seq-bwt-eins}, 
the structure RL is the most 
time-efficient up to 20 millions of sequential operations, and the most
energy-efficient up to 25 millions of operations. Both the running time and the
energy consumption depend directly on the number of instructions, CPU cycles, cache
accesses to the different levels of the memory hierarchy, among other
factors. The graph of the 
number of instructions in Figure~\ref{fig:seq-bwt-eins} shows that the \civ{}s
perform more instructions than the plain vector. In
general, \civ{}s perform several internal operations in order to recover an
element of the compacted vector, thus, increasing the number of machine
instructions. However, the instructions performed by the \civ{}s need less CPU
cycles to be completed than the instructions of the plain vector, as it is shown
in the graph of CPU cycles per instruction 
of Figure~\ref{fig:seq-bwt-eins}. When the number of operations is zero, the CPU
cycles per instruction is high since the latency involved in the loading of the
\civ{}s is also high. When the number of operations increases, the cost of
loading the data structures is amortized, reducing the average number of cycles
per instruction. The fact that \civ{}s need less CPU cycles per instruction than
the plain vector, suggests that the memory accesses involved
in some instructions are solved in lower levels of the memory hierarchy.
It can be corroborated in the graphs of L1d and L3 cache loads
of Figure~\ref{fig:seq-bwt-eins}, where \civ{}s use intensively the
L1d cache compared to the plain vector, meanwhile most of the \civ{}s have
less accesses to the L3 cache in comparison with the plain vector.
The size of the cache memories and their set-associativity have an impact on the
running time and the energy consumption of running applications. Cache memories
with reduced size and small 
set-associativity are more time- and energy-efficient than larger
memories~\cite{Hennessy:2017:CAS:3207796}. 
Accordingly, the L1d cache memory of the machine used in our experiments
is more time- and energy-efficient than the L3 cache memory, since the L1d cache
has a size of 32 KB and is 8-way set-associative, meanwhile the L3 cache has a
size of 20 MB and is 20-way set-associative. Unlike the cache behavior of the plain vector, \civ{}s exhibit a higher rate of memory accesses to L1d cache than to L3 cache, which may
explain that for more than 25 millions of sequential operations, the plain
vector is most time- and energy-efficient. 
This analysis is also valid for the {\tt psi} vector of the dataset {\tt kernel}
(Figure~\ref{fig:seq-kernel-psi}) and the {\tt lcp} vector of the dataset {\tt
  dblp} (Figure~\ref{fig:seq-dblp-lcp}).
For the {\tt psi} vector of {\tt kernel}, the most time- and energy-efficient
\civ{} is {\tt s9-zz}, and for the {\tt lcp} vector of {\tt dblp} is {\tt s9}.

From the results we observe that there is a strong correlation between the
energy consumption to load the \civ{}s in memory and the size of the
\civ{}s. However, when the  operations are performed, there is not a clear
correlation between energy consumption and size. For instance, for the {\tt lcp} vector of the {\tt dblp} dataset, the \civ{} {\tt
  s9} is the eighth smallest compact vector and the most efficient from an energy point of view. A
similar situation occurs with the {\tt psi} and {\tt bwt} vectors.

Regarding the relation between running time and energy consumption, there
exists a strong correlation: the faster the \civ{}, the more energy efficient.
Nevertheless, in some particular situations, the relation is broken. Such
situations are marked with vertical lines in
Figures~\ref{fig:sequential_access_1} and \ref{fig:sequential_access_2}. For instance, in
Figure~\ref{fig:seq-dblp-lcp}, in the range of 26 to 28 millions of
operations, the plain vector is faster than the data structure {\tt s9-zz}, but
{\tt s9-zz} is more energy-efficient. We leave as future work a more complete study
of such particular situations, in order to design energy-efficient \civ{}s.

Figures~\ref{fig:random_access_1} and \ref{fig:random_access_2} show the random access pattern. As expected, the
random access pattern takes more time and energy than the sequential access
pattern, since the former does not have the spatial data locality of the
latter. Moreover, for some of the techniques, each time that an element is accessed randomly,
several values must be decoded up to the desired element, starting from the closest sampled value.
The previous explanation of the impact of memory accesses, instructions
and CPU cycles per instruction on the running time and energy consumption of the \civ{}s
remains valid for the random access scenario. Among all the tested \civ{}s,
{\tt dac} and {\tt pfd} exhibit a remarkable improvement compared to their
behavior in sequential access scenario. In particular, both {\tt dac} and {\tt
  pfd} improve their behavior in the number of instructions and the number of
L1d cache accesses, thus, reducing their relative energy consumption.

From the energy point of view, the compact vectors storing differences are more
energy-efficient for sequential access pattern, performing worse for the random
access pattern.
This is an expected
behaviour, as retrieving the original value at a random position of a \civ{} storing differences requires
the use of sampling and the decompression of several positions instead of just one, thus,
nullifying the benefits of its fast direct access.

In summary, our experimental study provides evidence than the \civ{}s are a valid energy-efficient alternative
to manipulate integer vectors. For an increasing number of accesses to the elements of a vector, the
\civ{}s reduce their energy-efficiency. Therefore, the number of operations over the \civ{}s is a factor
that must be considered. For future work, this finding will allow us to define
favorable scenarios for the usage of \civ{}s in the design of energy-efficient algorithms.

Our experiments on access patterns revealed that such patterns have an important impact
in the energy behavior of the \civ{}s. If for some applications we could know in
advance the most common access patterns, we could select the most suitable \civ{} for
such application, potentially reducing its overall energy consumption. Otherwise,
our experiments suggest that \civ{}s such as {\tt dac} and {\tt s9} are a better
alternative, since their energy behavior is less affected with the access patterns.
Such \civ{}s, {\tt dac} and {\tt s9}, could be a good starting point to design energy-efficient
compact vectors for general scenarios.

\section{Conclusions and Future Work}

In this work we studied the energy behavior of several compact integer vectors for typical operations.
We performed experiments with different datasets, measuring metrics such as time, energy consumption,
number of instructions, memory transfers, among others. Our results suggest that compact integer vectors
offer a good alternative for the energy-efficient manipulation of vectors. This work is the first one providing
evidence that compact data structures may be considered in the design of energy-efficient algorithms.

As future work, we plan to develop a more low-level framework to better characterize all the factors
that impact on the energy consumption. Not only the computer architecture, memory hierarchy and memory
access patterns may be taken into account, but other factors, such as temperature inside and outside
the computation device, the complexity of the instructions used, among others. This will allow us to
propose new energy models that will make possible the design of algorithms that consider energy during
their design step.

\section*{Acknowledgment}

The authors thank Gonzalo Navarro for suggesting this line of research some time ago, and finally creating the link between the authors that made possible this collaboration.

\bibliographystyle{IEEEtran}
\bibliography{refs}

\begin{IEEEbiography}[{\includegraphics[width=1in,height=1.25in,clip,keepaspectratio]{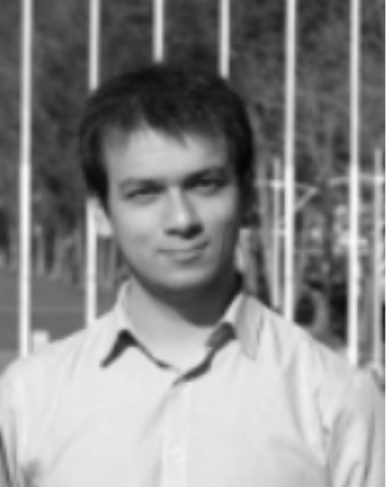}}]{Jos{\'e} Fuentes-Sep\'ulveda} is a Postdoctoral Researcher at University of Chile under the supervision of Gonzalo Navarro, PhD. He received his degree of Doctor in Computer Science from the University of Concepci\'on in 2016. His research interests include the design and analysis of algorithms and data structures, data compression and parallel algorithms.
\end{IEEEbiography}

\begin{IEEEbiography}[{\includegraphics[width=1in,height=1.25in,clip,keepaspectratio]{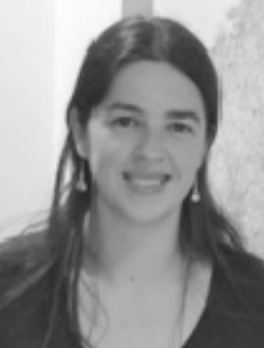}}]{Susana Ladra} is Associate Professor at the University of A Coru\~na, where she obtained her degree in Computer Science Engineering in 2007 and her Ph.D. degree in Computer Science in 2011 at the same university. She also received her Bachelor in Mathematics from the National Distance Education University (UNED) in 2014. Her fields of interests include the design and analysis of algorithms and data structures, data compression and data mining in the fields of information retrieval and bioinformatics. She has published more than 40 papers in various international journals and conferences and is principal investigator of several national and international research projects.
\end{IEEEbiography}

\EOD

\end{document}